\documentclass[journal=jctcce,manuscript=article]{achemso}   %Journal of Chemical Theory & Computation => journal=jctcce
\usepackage[version=3]{mhchem} % Formula subscripts using \ce{}
\usepackage{multirow}
\usepackage{soul}
\usepackage{array}
\usepackage[usenames,dvipsnames]{color} 
\usepackage[utf8]{inputenc}
\usepackage[T1]{fontenc}
\PassOptionsToPackage{hyphens}{url}
\usepackage{hyperref,url}
\usepackage{mathtools}
\usepackage{bm}
\usepackage{braket}
\usepackage{calrsfs}
\usepackage{multicol}
\usepackage{multirow}
\usepackage{makecell} 
\definecolor{mygreen}{rgb}{0.2,0.7,0.2}
\usepackage{dcolumn}
\usepackage{amsmath}
\usepackage{latexsym}
\usepackage{subcaption}
\usepackage{graphicx}
\usepackage{amssymb}
\usepackage{lineno} 
\usepackage{float}
\usepackage{physics}
\DeclareMathAlphabet{\mathcal}{OMS}{zplm}{m}{n}
\usepackage{bbm}

\definecolor{mygreen}{rgb}{0.2,0.7,0.2}

\renewcommand{\arraystretch}{0.6} % Space between pmatrix rows (1 standard)

\newcommand*{\LightComments}{}%

\ifdefined\NoComments
 
\else
 \ifdefined\LightComments
  
\else
  
 \fi
\fi
\author{A. Garc\'ia-Bl\'azquez}
\email{antonio.garcia@uni-wuerzburg.de}
\affiliation{~Institut f\"ur Physikalische und Theoretische Chemie, Universit\"at W\"urzburg, 97074 W\"urzburg, Germany}
\alsoaffiliation{Departamento de F\'isica de la Materia Condensada, Universidad Aut\'onoma de Madrid, 28049 Madrid, Spain}

\author{A. Erba}
\affiliation{~Dipartimento di Chimica, Universit\`a
di Torino, 10125 Torino, Italy}

\author{A. Boccuni}
\affiliation{~Dipartimento di Chimica, Universit\`a
di Torino, 10125 Torino, Italy}

\author{J.K. Desmarais}
\affiliation{~Dipartimento di Chimica, Universit\`a
di Torino, 10125 Torino, Italy}

\title{Damped polarizability and first hyperpolarizability of large systems with hybrid functionals: self-consistent and Lanczos solvers}
\begin{document}
\graphicspath{{Images/}}

%\begin{tocentry}
%\includegraphics[width=8.25cm,height=4.45cm]{ToC.png}
%\end{tocentry}


\begin{abstract} 
We present a damped coupled-perturbed Kohn-Sham method to compute the dynamical polarizability and first hyperpolarizability of finite systems, implemented in the \texttt{CRYSTAL} code for semi-local, global and range-separated hybrid functionals. The approach is based on the adiabatic time-dependent density functional theory for the complex-valued (hyper)polarizabilities, describing optical phenomena such as absorption, and the nonlinear second-harmonic generation, rectification and Pockels effect. These are obtained either at individual frequencies with a self-consistent solver based on Pulay extrapolation, or at an arbitrary grid of frequency and damping factors with a Lanczos solver, from which the first-order amplitudes required in the hyperpolarizability can be reconstructed. With hybrid functionals, the self-consistent solver tends to become impractical at some point beyond the absorption onset, where there are many excitations for which the external frequency lies between their energy and those of their contributing single-particle transitions, and its convergence eventually stalls for large systems. In contrast, the convergence of the Lanczos method is essentially independent of size, hence the cost of the dynamical response calculation scales with system size as the hybrid ground-state calculation. The implementation is validated in a small molecule and, as an application, we study the first and second-order optical response of a series of hexagonal boron nitride flakes of up to 324 atoms. The use of range-separated hybrid functionals allows to capture the strongly-bound first excitation of the monolayer, and evidences the failure of the self-consistent method to converge excitations with strong electron-hole attraction. 
\end{abstract}

\flushbottom
\maketitle

\thispagestyle{empty}

\section{Introduction}
Time-dependent density functional theory (TDDFT) constitutes one of the most important methods to determine the neutral excitations of systems with many electrons, owing to its balance between accuracy and computational cost \cite{RungeGross1984,MarquesGross2004,UllrichTDDFT,CasidaHuixRotllant2012,Maitra2016}. These excited states are directly probed by external electromagnetic radiation, determining the optical response of the system. Numerous implementations are nowadays available in electronic structure codes, although the most common approaches consist in either resolving the dynamics in real time over femtosecond scales (RT-TDDFT) \cite{Yabana1996_RT,Lopata2011_NWChemRT,Li2020_RTchemrev,FHIaims_RT,PySCF_RT}, or solving the Casida equation for the excitation energies \cite{Casida1995}, either by full diagonalization or for a limited number of them with iterative eigensolvers \cite{Bauernschmitt1996_excit,Stratmann1998,Furche2016,vecharynski2017efficient}.
RT-TDDFT yields the whole linear spectrum from a single propagation per field direction, which makes it competitive for large systems \cite{Tussupbayev2015_RTvsLR} although achieving fine resolution might turn expensive, and the nonlinear responses require multiple such processes per frequency with different field strengths \cite{Takimoto2007_RTbeta,Ding2013_RTbeta}. In the Casida formulation, on the other hand, computing the spectra of large systems over a large frequency window is computationally demanding since the density of excited states grows with system size, which has motivated approximate scalable schemes \cite{Grimme2013_sTDA,Bannwarth2014_sTDDFT}.

An alternative approach is to solve the linear-response equations directly in frequency space, for a perturbation of given frequency \cite{Gisbergen1998,Agren2002_beta,Octopus2007,NWCHEM2011,Kummel2018}. This formulation, in terms of the first-order response of the orbitals, is traditionally known as coupled-perturbed Kohn-Sham (CPKS) or Hartree-Fock (CPHF) \cite{hurst1988ab,KarnaDupuis1991,ferrero2008_1}, otherwise known as adiabatic time-dependent density functional perturbation theory (TD-DFPT) \cite{Baroni2001_DFPT,Walker2006_Lanczos} particularly amenable for application at higher orders. %, not necessarily employing the Sternheimer formulation to factor out the virtual states. 
It bypasses the time propagation and the calculation of individual excited states, and its equations have often been solved self-consistently for each frequency, which may be the most convenient choice if only a few of them are required, for instance for static calculations or comparison with experiments at specific wavelengths. In order to obtain the absorption, and any type of optical response close to the resonances at excitation energies, the frequency must be extended to the complex plane including a damping factor, which physically represents a constant relaxation rate of the excited states \cite{Norman2001_CPP,Norman2011_review,Jensen2005_damped,kristensen2009quasienergy,DALTON2017,ADF2016}.

A different formulation is provided by Krylov-subspace recursions built with the frequency-independent response operator, which yields the response at all frequencies from a single recursion per field direction. In the Liouville-Lanczos approach to TD-DFPT, the dynamical polarizability is obtained as a continued fraction of the Lanczos coefficients, which can be evaluated at any complex frequency once a sufficiently long chain of such coefficients has been built \cite{Walker2006_Lanczos,C60_calc_Baroni,Malcioglu2011_turboTDDFT}. Since the response operator is not Hermitian, these chains were initially built with the non-symmetric Lanczos algorithm. A pseudo-hermitian form of the recursion, which exploits the positive definiteness of the electronic Hessian, requires half the operator builds and does not suffer from the breakdowns of the non-symmetric formulation \cite{Gruning2011_pseudoHermitianLanczos,Ge2014_turboTDDFT}. The pseudo-hermitian formulation can analogously be recast into a symmetric form, in which the recursion acts on a product of two matrices of half the dimension of the response operator \cite{Brabec2015,Shao2018_SPLanczos}. These variants of the Lanczos method provide an efficient and numerically stable approach to solve for the dynamic polarizability although, to our knowledge, they have so far not been used to obtain hyperpolarizabilities.

The first hyperpolarizability $\beta(\omega,\omega')$ is a function of two frequencies whose subdomains correspond to different second-order optical processes, such as second-harmonic generation (SHG), the electro-optic Pockels effect (POC) and optical rectification (REC) \cite{KarnaDupuis1991,Willetts1992_conventions,BoydNLO}. The difference between the last two arises as a consequence of the damping. 
As in the undamped case, the damped $\beta$ can be obtained from first-order solutions alone through the $(2n+1)-$rule \cite{norman2005nonlinear,ADF2014}, involving the derivative of the exchange-correlation kernel of one order higher. We derive rigorously the expressions of the (hyper)polarizabilities for our formulation of the damping, with a complete discussion on the Lagrange multipliers in the damped case. This first hyperpolarizability is typically reported for small or medium-sized molecules and small metal clusters \cite{norman2005nonlinear,ADF2014,DALTON2017,Ahmadzadeh2022_SHG}, while for larger systems $\beta$ has usually been computed in the static limit \cite{knoppe2018role} or without damping using simplified Hamiltonians \cite{Loeffelsender2024_WIREs}. 
Its value depends strongly on the amount and range of Fock-exchange \cite{Kamiya2005_LCCPKS,Lescos2020_RSHbeta}, which indicates the importance of an efficient treatment of the response with range-separated hybrid functionals.

The accuracy of the TDDFT description of the optical response is ultimately determined by the exchange-correlation (XC) kernel. The adiabatic kernels of semi-local functionals lack the extreme non-locality of the exact kernel (with an exception being represented by the treatment of Ref. \cite{nazarov2011optics}) which is essential for charge-transfer excitations in molecules \cite{Baer2010_tunedRSH} and for excitonic effects in extended systems \cite{RefaelyAbramson2015,Ohad2023_TDWOTSRSH}. Hybrid functionals may recover the exact long-range behavior through the non-local Fock-exchange, and range-separated hybrids, which include the full Fock-exchange at long range, can be tuned to fulfill exact conditions in finite systems \cite{Baer2010_tunedRSH,Stein2010_OTRSH,Korzdorfer2011_omega}. Screened variants, in which the long-range fraction of Fock exchange is set by the dielectric screening, extend this approach to solids \cite{Wing2021_WOTSRSH,Ohad2023_TDWOTSRSH}, including 2D materials \cite{Ramasubramaniam2019_SRSH,Camarasa2023_SRSH,Camarasa2024_WOTSRSH}, with results sometimes comparable to the many-body Bethe-Salpeter equation (BSE) at a lower cost as well as to the experiment \cite{Ramasubramaniam2019_SRSH,Camarasa2023_SRSH,Camarasa2024_WOTSRSH}. Particularly challenging systems are those with strongly bound excitons such as hexagonal boron nitride (hBN), a prototypical 2D insulator whose optical properties are dominated by these states with strong electron-hole attraction \cite{galvani2016excitons,Elias2019_monolayerBN,zhang2022intervalley}.

The present work describes the damped self-consistent methodology newly implemented in the \texttt{CRYSTAL} code \cite{crystal17,crystal23}, at present for finite systems.  It incorporates global and range-separated hybrid functionals at a relatively low cost \cite{pisani1980exact,causa1988treatment,pisani1988,dovesi2026hartree,ambrogio2026efficient}, for closed-shell and collinear open-shell systems, and employs the unperturbed point group symmetry to accelerate the calculation of the bi-electronic matrix elements (which in practice constitute the main computational bottleneck).\cite{thesis_jacopo}  We describe two solvers for the first-order equations: a self-consistent one with the direct inversion of the iterative subspace (DIIS) \cite{Pulay1980}, applied separately at each frequency, and a Lanczos recursion that provides the response at all frequencies and damping factors from a single chain per field direction, from which we furthermore obtain the first hyperpolarizability. As a case study, we compute and analyze the first and second-order responses in hBN flakes of increasing size employing range-separated hybrid functionals, comparing the absorption with the BSE result in the periodic limit. The scaling and convergence of the Lanczos and self-consistent solvers are analyzed side by side, discussing the role of Fock-exchange in the convergence deterioration of the latter solver for large systems beyond the absorption onset.

\section{Theory}
\subsection{The unperturbed system}
Atomic units are employed throughout the text. $i,j$ and $a,b$ label occupied and virtual states, respectively, and $u,v,w$ label Cartesian directions. Consider a molecular or cluster system described in the absence of external fields by the one-electron wavefunctions
%\begin{equation}\label{psi0}
$\psi_{n\sigma}^{(0)}(\bm{r})=\sum_{\mu}c_{\mu,n}^{\sigma\;(0)}\varphi_{\mu}(\bm{r})$,
%\end{equation}
where $\varphi_{\mu}$ are atomic orbitals (AOs), in the present work of Gaussian type, $\sigma=\uparrow,\downarrow$ labels the (collinear) spin projection referred to a given quantization axis, and the molecular orbital (MO) coefficients $c_{\mu,n}^{\sigma\;(0)}$ and single-particle energies $\varepsilon_{n}^{(0)}$ are obtained in the usual way by solving the generalized Kohn-Sham (GKS) equations in the AO basis. The latter are here built in general from a hybrid DFT Hamiltonian
%\begin{equation}\label{Hmat0}
$H_{\mu,\nu}^{\sigma\;(0)}=h_{\mu,\nu}^{(0)}+
B_{\mu,\nu}^{\sigma\;(0)} + V^{\sigma\;(0)}_{\mu,\nu} $, 
%\end{equation}
where the superindex $0$ indicates the absence of external fields, 
$h$ contains the mono-electronic integrals (possibly including atomic pseudo-potentials), and 
\begin{itemize}
\item $B$ contains the bi-electronic Hartree and Fock-exchange integrals, 
\begin{equation}\label{Bmat0}
B_{\mu,\nu}^{\sigma\;(0)}= \sum_{\mu',\nu'}\sum_{\sigma'}P_{\mu',\nu'}^{\sigma'\;(0)}\left[(\mu\nu\vert\mu'\nu') -\delta_{\sigma,\sigma'}(\mu\mu'\vert\nu\nu')\right]
\end{equation}
Here we have introduced the (spin-resolved) AO density matrix
$P_{\mu,\nu}^{\sigma\;(0)}=\sum_{i}c^{\sigma\;(0)}_{\mu,i}c^{\sigma\;(0)*}_{\nu,i}$, 
written more compactly as $P^{\sigma(0)}=c^{\sigma(0)}N^{\sigma}c^{\sigma(0)\dagger}$ ($N^{\sigma}$ being the diagonal matrix of occupation numbers), as well as the shorthand notation \newline
$(\mu\nu\vert\mu'\nu')=\int\varphi_{\mu}(\bm{r})\varphi_{\nu}(\bm{r})\frac{1}{\abs{\bm{r}-\bm{r}'}}\varphi_{\mu'}(\bm{r}')\varphi_{\nu'}(\bm{r}')d\bm{r}d\bm{r}'$.
\item $V^\sigma$ is the semi-local part of the exchange-correlation (XC) potential 
\begin{equation}\label{XCmat0}
V^{\sigma\;(0)}_{\mu,\nu} =\int\varphi_{\mu}(\bm{r})\varphi_{\nu}(\bm{r}) \left. \frac{\delta E_{\text{xc}}[n_\uparrow,n_\downarrow]}{\delta n_{\sigma}(\bm{r})} \right\vert_{n_\uparrow^{(0)},n_\downarrow^{(0)}} d\bm{r}
\end{equation}
The functional derivative of the XC energy $E_{\text{xc}}[n_\uparrow,n_\downarrow]$ is evaluated at the unperturbed (i.e. zero field) spin density 
$n_{\sigma}^{(0)}(\bm{r})=
\sum_{\mu,\nu}P_{\mu,\nu}^{\sigma\;(0)}\varphi_{\mu}(\bm{r})\varphi_{\nu}(\bm{r})$. 
\end{itemize}
The XC/Fock-exchange mixing parameter as well as a possible range-separation have been omitted for simplicity in the notation: for a hybrid or range-separated functional, the exchange integrals in $B$ are scaled by the fraction of exact exchange, or replaced by those of the long-range (or short-range) Coulomb operator, while $E_{\text{xc}}$ contains the correlation and the complementary fraction of (semi-local) exchange. For the unperturbed problem we consider current-less collinear magnetic systems (in particular no spin-orbit coupling), such that $c_{\mu,n}^{\sigma\;(0)}$ are real-valued orbital coefficients.

\subsection{The time-dependent equations}
Under an external electromagnetic radiation defined by the transversal field $\bm{E}(\bm{r},t)$, the electronic Hamiltonian acquires a dynamic interaction or coupling term $H^{I}$. In the long-wavelength approximation, based on the assumption that the field wavelength is much larger than the dimension of the system (which is reasonable for any nanoscopic system under ultraviolet or visible radiation), the spatial variation of the field is neglected in the interaction. Then, the electric field is expanded in Fourier components as

\begin{equation}\label{Efield}
\bm{E}(t)=\sum_{l}\left[\bm{E}(\omega_{l})e^{-i\omega_{l} t} + \bm{E}(-\omega_{l})e^{i\omega_{l} t}\right] 
\end{equation}
where $\bm{E}(-\omega)=\bm{E}(\omega)^{*}$ [$\bm{E}(\omega)$ being real for linearly polarized light]. 
Throughout the document we shall denote an arbitrary frequency, of any sign, by $\omega$.

The coupling Hamiltonian in the long-wavelength approximation can be written in the length gauge as
$H^{I}_{\mu,\nu}(t)=-\bm{\mu}_{\mu,\nu}\cdot\bm{E}(t)$,  
in terms of the electric dipole integrals (for a finite system and including the electron charge) $\mu^{u}_{\mu,\nu}=-\int\varphi_{\mu}(\bm{r})r^{u}\varphi_{\nu}(\bm{r})d\bm{r}$, $u=x,y,z$, which we treat as a perturbation to the ground-state. 
Besides appearing as $H^{I}$ in the Hamiltonian, the perturbation changes the density, and through it, the self-consistent Hartree, exchange and correlation potentials.
In particular,
\begin{equation}\label{Pdensity}
P_{\mu,\nu}^{\sigma}(t)=\sum_{i}c^{\sigma}_{\mu,i}(t)c^{\sigma}_{\nu,i}(t)^{*},
\end{equation}
\begin{equation}\label{density}
n_\sigma(\bm{r},t)=\sum_{\mu,\nu}P_{\mu,\nu}^{\sigma}(t)\varphi_{\mu}(\bm{r})\varphi_{\nu}(\bm{r})
\end{equation}
and likewise for the other perturbed quantities in terms of $P^{\sigma}(t)$.   

For the XC potential we employ the adiabatic approximation, by which the same functional as in the static theory is evaluated at the instantaneous spin densities $n_\sigma(\mathbf{r},t)$ (i.e. neglecting any memory, or retardation effects), that is,

\begin{equation}\label{XCmat}
V^{\sigma}_{\mu,\nu}(t) \approx
\int\varphi_{\mu}(\bm{r})\varphi_{\nu}(\bm{r})\left.\frac{\delta E_{\text{xc}}[n_\uparrow,n_\downarrow]}{\delta n_{\sigma}(\bm{r})}\right\vert_{n_\uparrow(t),n_\downarrow(t)} d\bm{r}
\end{equation}
The full perturbed Hamiltonian then reads 
\begin{equation}\label{Ht}
H^{\sigma}_{\mu,\nu}(t)=h^{(0)}_{\mu,\nu}+ B^{\sigma}_{\mu,\nu}(t) + V^{\sigma}_{\mu,\nu}(t) + H^{I}_{\mu,\nu}(t)
\end{equation}

\textbf{The damped TDGKS equations.}
In terms of \eqref{Ht}, the orbital coefficients obey the adiabatic time-dependent generalized Kohn-Sham (TDGKS) equations of collinear spin DFT \cite{RungeGross1984,Seidl1996_GKS}, to which we add a relaxation term proportional to $\eta$ \cite{norman2005nonlinear,ADF2014} (see discussion below). This way, the TDGKS can be written in the AO basis as
\begin{equation}\label{TDKS}
H^{\sigma}(t)c^{\sigma}(t)-iS\frac{d c^{\sigma}(t)}{dt}-i\eta S[c^{\sigma}(t)-c^{\sigma\:(0)}]=Sc^{\sigma}(t)\lambda^{\sigma}(t)
\end{equation}
where $S$ is the overlap matrix, $\lambda$ is a matrix of Lagrange multipliers (in the MO basis) associated to the orthonormality constraint 
%\begin{equation}\label{orthonorm}
$\int\psi_{m}^{*}(\bm{r},t)\psi_{n}(\bm{r},t)d\bm{r}=\delta_{m,n}$ $\forall t$,
%\end{equation}
and $\eta>0$ is the rate of relaxation, also known as damping parameter.
Since \eqref{TDKS} is diagonal in spin space for our collinear open shell configuration, we shall drop the spin indices in the instances where it is trivial. 

The effect of the relaxation term is most transparent in the density matrix. Writing \eqref{Pdensity} as $P(t)=c(t)Nc(t)^{\dagger}$, taking the time derivative and inserting \eqref{TDKS} yields
\begin{equation}\label{dPdt}
\frac{dP}{dt}=-i\left(S^{-1}HP-PHS^{-1}\right)-\eta\left[\left(c-c^{(0)}\right)Nc^{\dagger}+cN\left(c-c^{(0)}\right)^{\dagger}\right]+ic\left(\lambda N-N\lambda^{\dagger}\right)c^{\dagger}
\end{equation}
In the right-hand side, the first term is the analogue of the von Neumann equation $i\,d\hat{\rho}/dt=[\hat{H},\hat{\rho}]$ for the one-particle density operator $\hat{\rho}=\sum_{\mu,\nu}\ket{\varphi_{\mu}}P_{\mu,\nu}\bra{\varphi_{\nu}}$, written in the non-orthogonal AO basis, i.e. the evolution generated by the Hamiltonian. The second term can be written as $-\eta[P-P^{(0)}+(c-c^{(0)})N(c-c^{(0)})^{\dagger}]$, which to first order in the field, see \eqref{Cexpans}, includes only contributions from $-\eta[P-P^{(0)}]$; hence the perturbed density matrix relaxes to the unperturbed one (plus a correction of second-order) at a rate $\eta$, or equivalently $\eta$ is an inverse lifetime, common to all excitations. The third term in the right-hand side of \eqref{dPdt} contains the Lagrange multipliers, which must fulfill the relation 
\begin{equation}\label{lambdaC}
\lambda-\lambda^{\dagger}=-i\eta(c-c^{(0)})^{\dagger}S(c-c^{(0)}) 
\end{equation}
obtained by taking the time derivative of the orthonormality constraint and inserting \eqref{TDKS}. This fixes the anti-hermitian part of $\lambda$, which is zero in the absence of damping, and second-order in the field for $\eta\neq0$. 
On the other hand, the last term in \eqref{dPdt} depends only on $\lambda_{\text{occ,occ}}-\lambda_{\text{occ,occ}}^{\dagger}$ and $\lambda_{\text{virt,occ}}$. The former is fixed by \eqref{lambdaC} and gauge-invariant, whereas the latter constitutes an arbitrary transformation of the occupied into the virtual subspace that directly changes the evolution of the density matrix (and hence physical observables over time). Therefore, $\lambda_{\text{virt,occ}}$ must vanish, and the off-diagonal blocks of \eqref{lambdaC} become a restriction on $\lambda_{\text{occ,virt}}$, in general nonzero due to the damping factor.

\textbf{Perturbative equations in frequency space.}
The TDGKS equations can be solved by expanding the MO coefficients in powers of the field amplitudes,
\begin{equation}\label{Cexpans}
c(t)=c^{(0)}+\sum_{u}\sum_{\omega}c^{u}(\omega)E^{u}(\omega)e^{-i\omega t}
+\frac{1}{2}\sum_{u,v}\sum_{\omega,\omega'}c^{u,v}(\omega,\omega')E^{u}(\omega)E^{v}(\omega')e^{-i(\omega+\omega')t}+\mathcal{O}(E^{3})
\end{equation}
where $\omega,\omega'$ span the frequencies $\pm\omega_{l}$ of \eqref{Efield}, and $c^{u,v}(\omega,\omega')=c^{v,u}(\omega',\omega)$. The expansion \eqref{Cexpans} defines a steady state decomposed in the frequencies of the field and their sums. The terms of third and higher order will be disregarded throughout. 

The expansion \eqref{Cexpans} translates into an analogous one for the other time dependent quantities. 
For the AO density matrix, inserting \eqref{Cexpans} in \eqref{Pdensity} and grouping terms proportional to the same field amplitudes, one finds for the first and second-order frequency coefficients
\begin{equation}\label{P1}
P^{u}(\omega)=c^{u}(\omega)Nc^{(0)\dagger} + c^{(0)}Nc^{u}(-\omega)^{\dagger}
\end{equation}
\begin{equation}\label{P2}
P^{u,v}(\omega,\omega')=c^{u,v}(\omega,\omega')Nc^{(0)\dagger} + c^{(0)}Nc^{u,v}(-\omega,-\omega')^{\dagger} + c^{u}(\omega)Nc^{v}(-\omega')^{\dagger} + c^{v}(\omega')Nc^{u}(-\omega)^{\dagger}
\end{equation}

For the spin density \eqref{density}, one can identify
\begin{equation}\label{n1}
n^{u}_\sigma(\bm{r},\omega)=\left.\frac{\partial n_\sigma}{\partial E^{u}(\omega)}\right\vert_{\bm{E}=\bm{0}} =
\sum_{\mu,\nu}P_{\mu,\nu}^{\sigma\;u}(\omega)\varphi_{\mu}(\bm{r})\varphi_{\nu}(\bm{r})
\end{equation}
\begin{equation}\label{n2}
n^{u,v}_\sigma(\bm{r},\omega,\omega')=\left.\frac{\partial^{2}n_\sigma}{\partial E^{u}(\omega)\partial E^{v}(\omega')}\right\vert_{\bm{E}=\bm{0}} =
\sum_{\mu,\nu}P_{\mu,\nu}^{\sigma\;u,v}(\omega,\omega')\varphi_{\mu}(\bm{r})\varphi_{\nu}(\bm{r})
\end{equation}
Consequently, an expression for the frequency coefficients of the full Hamiltonian in terms of these quantities can be obtained from \eqref{Ht}, yielding at first order 
\begin{equation}\label{H1}
H_{\mu,\nu}^{\sigma\;u}(\omega)=
-\mu^{u}_{\mu,\nu}+
B_{\mu,\nu}^{\sigma\;u}(\omega)
+V_{\mu,\nu}^{\sigma\;u}(\omega)
\end{equation}
\begin{equation}\label{B1}
B_{\mu,\nu}^{\sigma\;u}(\omega)=
\sum_{\mu',\nu'}\sum_{\sigma'}P^{\sigma'\;u}_{\mu',\nu'}(\omega)\left[(\mu\nu\vert\mu'\nu') -\delta_{\sigma,\sigma'}(\mu\mu'\vert\nu\nu')\right]
\end{equation}
\begin{equation}\label{V1}
V_{\mu,\nu}^{\sigma\;u}(\omega)=\sum_{\sigma'}\int\varphi_{\mu}(\bm{r})\varphi_{\nu}(\bm{r})\left.f_{\text{xc}}^{\sigma,\sigma'}(\bm{r},\bm{r}')\right\vert_{n^{(0)}_\uparrow,n^{(0)}_\downarrow}n^{u}_{\sigma'}(\bm{r}',\omega)d\bm{r}d\bm{r}'
\end{equation}
in terms of the (adiabatic) XC kernel
\begin{equation}\label{fxc}
f_{\text{xc}}^{\sigma,\sigma'}(\bm{r},\bm{r}')=\frac{\delta^{2}E_{\text{xc}}[n_\uparrow,n_\downarrow]}{\delta n_{\sigma}(\bm{r})\delta n_{\sigma'}(\bm{r}')}
\end{equation}
evaluated at the unperturbed spin densities owing to the zero-field condition in \eqref{n1}.
Likewise, employing 
the first equality in \eqref{n2}
it follows that the second-order coefficients are
\begin{equation}\label{H2}
H_{\mu,\nu}^{\sigma\;u,v}(\omega,\omega')=
B_{\mu,\nu}^{\sigma\;u,v}(\omega,\omega')
+V_{\mu,\nu}^{\sigma\;u,v}(\omega,\omega')
\end{equation}
\begin{equation}\label{B2}
B_{\mu,\nu}^{\sigma\;u,v}(\omega,\omega')=
\sum_{\mu',\nu'}\sum_{\sigma'}P^{\sigma'\;u,v}_{\mu',\nu'}(\omega,\omega')\left[(\mu\nu\vert\mu'\nu') -\delta_{\sigma,\sigma'}(\mu\mu'\vert\nu\nu')\right]
\end{equation}
\begin{equation}\label{V2}
\begin{aligned}
 V_{\mu,\nu}^{\sigma\;u,v}(\omega,\omega') =
&\int\varphi_{\mu}(\bm{r})\varphi_{\nu}(\bm{r}) \sum_{\sigma^\prime} \left[
n^{u,v}_{\sigma^\prime}(\bm{r}',\omega,\omega') \left.f_{\text{xc}}^{\sigma,\sigma'}(\bm{r},\bm{r}')\right\vert_{n^{(0)}_\uparrow,n^{(0)}_\downarrow} + \right. \\
&\left. n^{u}_{\sigma^\prime}(\bm{r}',\omega)\int d\bm{r}'' \sum_{\sigma^{\prime \prime}} n^{v}_{\sigma^{\prime \prime}}(\bm{r}'',\omega')
 \left.\frac{\delta f_{\text{xc}}^{\sigma,\sigma'}(\bm{r},\bm{r}')}{\delta n_{\sigma^{\prime\prime }}(\bm{r}'')}\right\vert_{n^{(0)}_\uparrow,n^{(0)}_\downarrow} \right]
d\bm{r}d\bm{r}'
\end{aligned}
\end{equation}
In the LSDA, 
XC kernels reduce to partial derivatives of the local energy density, while the GGA expressions are more involved and can be found in \ref{AppC}. 

Note that, due to \eqref{P1} and \eqref{P2},
\begin{equation}\label{whermit}
P^{u}(-\omega)=P^{u}(\omega)^{\dagger}, \; P^{u,v}(-\omega,-\omega')=P^{u,v}(\omega,\omega')^{\dagger}
\end{equation}
as required by the hermiticity of $P(t)$. The same can be observed to hold for $n$ (where the matrix transpose is trivial) and $H$. Furthermore, $V$ and the Hartree term in $B$ are symmetric matrices at any frequency, unlike the Fock-exchange.

\subsection{The coupled-perturbed formulation}\label{secCP}
In frequency space, we look for a solution for the perturbed MO coefficients in the form
\begin{equation}\label{Udef}
c^{u}(\omega)=c^{(0)}U^{u}(\omega), \;\;
c^{u,v}(\omega,\omega')=c^{(0)}U^{u,v}(\omega,\omega')
\end{equation}
Inserting \eqref{Cexpans} and \eqref{Udef} in the orthonormality constraint, and grouping terms proportional to the same field amplitudes, one finds the usual 
$c^{(0)\dagger}Sc^{(0)}=I$
with $I$ the identity matrix, and consequently the following order-by-order orthonormality conditions
\begin{equation}\label{Uorth1}
U^{u}(-\omega)^{\dagger}=-U^{u}(\omega)
\end{equation}
\begin{equation}\label{Uorth2}
U^{u,v}(-\omega,-\omega')^{\dagger}+U^{u,v}(\omega,\omega')=
U^{u}(\omega)U^{v}(\omega')+
U^{v}(\omega')U^{u}(\omega)
\end{equation}
It can be observed that the orthonormality conditions do not completely determine $U$ order by order, in the same way as \eqref{lambdaC} does not completely determine $\lambda$, the remaining freedom representing a gauge choice. Once such choice has been made for $U$ in consistency with \eqref{Uorth1} or the corresponding higher-order condition, the Lagrange multipliers follow from the TDGKS equations and \eqref{lambdaC}, as discussed above with vanishing virt-occ block (at all orders, since it holds for the time expression). Equivalently, one may choose $\lambda$ satisfying \eqref{lambdaC} and determine $U$ from the TDGKS equations and the orthonormality conditions. 
We therefore expand $\lambda(t)$ as in \eqref{Cexpans}, aiming to find the most convenient multipliers at each order. For the zeroth order, it follows from \eqref{TDKS} 
$H^{(0)}c^{(0)} = Sc^{(0)}\lambda^{(0)}$
whereby $\lambda^{(0)}_{m,n}=\varepsilon^{(0)}_{m}\delta_{m,n}$ in compliance with the unperturbed GKS equations and the restriction \eqref{lambdaC}, which forces $\lambda^{(0)}$ to be hermitian. In the first order, the adiabatic TDGKS equations then lead to
\begin{equation}\label{TDKS1}
G^{u}(\omega) + \varepsilon^{(0)}U^{u}(\omega) - U^{u}(\omega)\varepsilon^{(0)} - (\omega+i\eta)U^{u}(\omega)= \lambda^{u}(\omega)
\end{equation}
where 
$G^{u}(\omega)=c^{(0)\dagger}H^{u}(\omega)c^{(0)}$.

Taking matrix elements in \eqref{TDKS1} it is observed that the diagonal of $U^{u}(\omega)$ would diverge at zero frequency in the absence of the damping factor $\eta$. Hence, for consistency with the case of a static perturbation (where $\eta$ is ignored) we choose a gauge in which the diagonal blocks of $U^{u}(\omega)$ are vanishing, which is known as the parallel-transport gauge \cite{Gonze1995,Jia2019_PTACE}. This can be done by setting $\lambda^{u}(\omega)=G^{u}(\omega)$ for the diagonal blocks and zero for the off-diagonal blocks, consistent with the restriction \eqref{lambdaC} at first order [note that the left-hand side at first order is $\lambda^{u}(\omega)-\lambda^{u}(-\omega)^{\dagger}$]. Then, from \eqref{TDKS1}, 
\begin{equation}\label{U1}
U^{u}_{m,n}(\omega)=\left\{\begin{aligned}&\frac{G^{u}_{m,n}(\omega)}{\omega+\varepsilon^{(0)}_{n}-\varepsilon^{(0)}_{m}+i\eta} , \;\;
&m\:\text{occ},n\:\text{virt} ; \text{ or } m\:\text{virt},n\:\text{occ} 
\\
&0, \;\;
&m\:\text{occ},n\:\text{occ} ; \text{ or } m\:\text{virt},n\:\text{virt} 
\end{aligned}\right.
\end{equation}
Note that the relative sign of the imaginary damping $i\eta$ in $U^{u}(\omega)$ and $U^{u}(-\omega)$ (as for other quantities) is fixed by the requirement of simultaneously satisfying the first-order adiabatic TDGKS equations, as well as the first-order orthonormality condition \eqref{Uorth1}. Higher-order $U$ matrices can be obtained analogously, see e.g. \eqref{U2} for the second order. The same can be done for the Lagrange multipliers, with the observation that  \eqref{lambdaC} at second order is
$\lambda^{u,v}(\omega,\omega')-\lambda^{u,v}(-\omega,-\omega')^{\dagger}= i\eta \left[U^{u}(\omega)U^{v}(\omega')+
U^{v}(\omega')U^{u}(\omega)\right]$, which by \eqref{U1} is block-diagonal and hence the second-order Lagrange multipliers, already having zero virt-occ block, must also be block-diagonal (which does not generally hold at higher orders for $\eta\neq0$).

These circular relations provide a method to compute $U$ self-consistently to first and second order. For $U^{u}(\omega)$, we start from the initial guess $H^{u}=-\mu^{u}$ for \eqref{H1} in $G^{u}(\omega)$ [equivalent to $P(t)=P^{(0)}$], compute the occ-virt and virt-occ blocks of $U^{u}(\omega)$ from \eqref{U1} [this includes $U^{u}(-\omega)$ via \eqref{Uorth1}], followed by $P^{u}(\omega)$ from \eqref{P1} and, after going through \eqref{n1}, \eqref{B1} and \eqref{V1}, a new $H^{u}(\omega)$ from \eqref{H1}. The cycle is iterated until $U^{u}(\omega)$ is converged, as monitored through the polarizability $\alpha^{uu}(\omega)$ \eqref{alpha} and possibly also the total change in $H^{u}(\omega)$, and the process is repeated for the 3 Cartesian directions unless not required due to symmetry of the selected optical tensors. 
On the other hand, the second-order $U^{u,v}(\omega,\omega')$ in \eqref{U2} can be computed with an additional self-consistent procedure, although in the present work we instead resort to the $(2n+1)-$rule to obtain the second-order response directly from $U^{u}(\omega)$. Alternatively, the first-order equations can be solved for all frequencies within a selected energy range with the Lanczos solver described in Section \ref{secLanczos}.

\subsection{Damped polarizability and first hyperpolarizability}\label{secAlphaBeta}
Our ultimate goal is to compute the dynamical polarizability $\alpha^{uv}(\omega)$ and first hyperpolarizability $\beta^{uvw}(\omega,\omega')$ of the system of many electrons, defined respectively as the first and second-order frequency coefficients [see \eqref{Cexpans}] of the electric dipole 
\begin{equation}\label{dipole}
d^{u}(t)=\int r^{u}[-n(\bm{r},t)+n_{\text{nuc}}(\bm{r})]d\bm{r}=
\sum_{\sigma}\sum_{\mu,\nu}\mu^{u}_{\mu,\nu}P^{\sigma}_{\mu,\nu}(t)
+\sum_{A}Z_{A}R^{u}_{A}
\end{equation}
where $Z_{A}$ and $\bm{R}_{A}$ are the nuclear charges and positions, in which we disregard dynamical effects.
Following the previous discussion we can write, in terms of the perturbed MO transformation matrices $U$ in \eqref{Udef}, and the matrix
\begin{equation}\label{sdip}
s^{u}=c^{(0)\dagger}\mu^{u}c^{(0)}
\end{equation}
the following expressions for the polarizabilities:
\begin{equation}\label{alpha}
\alpha^{uv}(\omega)=\Tr[\mu^{u}P^{v}(\omega)] =
\Tr\left\{N\left[
s^{u}U^{v}(\omega) - U^{v}(\omega)s^{u}
\right]\right\}
\end{equation}
\begin{equation}\label{beta_v0}\begin{aligned}
&\beta^{uvw}(\omega,\omega')=\Tr[\mu^{u}P^{v,w} (\omega,\omega')] = \\
&\Tr\left\{N\left[
U^{v,w}(-\omega,-\omega')^{\dagger}s^{u} + s^{u}U^{v,w}(\omega,\omega') 
+ U^{v}(-\omega)^{\dagger}s^{u}U^{w}(\omega')
+ U^{w}(-\omega')^{\dagger}s^{u}U^{v}(\omega)
\right]\right\}
\end{aligned}\end{equation}
where $\Tr$ denotes the trace in the indices of the corresponding matrix (either over AOs and spin states or over eigenstates, a distinction that is unimportant for a finite basis as long as it is clear which one applies to each matrix). The factor $1/2$ in \eqref{Cexpans} corresponds to the Taylor-series convention for $\beta$ \cite{Willetts1992_conventions}.
From \eqref{whermit} it follows that $\alpha^{uv}(-\omega)=\alpha^{uv}(\omega)^{*}$. Moreover, $\alpha^{uv}(\omega)=\alpha^{vu}(\omega)$, which can be seen more clearly from the inner product form \eqref{alphaLz}. With the convention \eqref{Efield}, $\Im\alpha^{uu}(\omega)>0$ for $\omega>0$, which corresponds to an absorption process, and the dipole strength function is computed as
%\begin{equation}\label{Sfun}
$S(\omega)=\frac{2}{\pi}\omega\Im[\overline{\alpha}(\omega)]$
%\end{equation} 
in terms of the isotropic polarizability, invariant under rotations,
\begin{equation}\label{alphaiso}
\overline{\alpha}(\omega)=\frac{1}{3}\left[\alpha^{xx}(\omega)+\alpha^{yy}(\omega)+\alpha^{zz}(\omega)\right]
\end{equation}

The second-order $U^{v,w}$ terms can be removed from \eqref{beta_v0}, as expected from the $(2n+1)-$rule, following the procedure of Refs. \cite{KarnaDupuis1991,Gisbergen1998}. As shown in \ref{AppA1}, the procedure can be extended to account for the complex damping, essentially because the second-order equation \eqref{TDKS2_pre} carries the same damping as the first-order \eqref{TDKS1} at $\omega+\omega'$. The final result reads
\begin{equation}\label{beta}\begin{aligned}
&\beta^{uvw}(\omega,\omega')= -\beta_{\text{xc}}^{uvw}(\omega,\omega') + 
\Tr\left\{N\left(\;
U^{u}(-\omega-\omega')^{*}[G^{v}(\omega),U^{w}(\omega')]_{-}  \:+
\right.\right. \\ 
&\left.\left. U^{v}(\omega)[G^{w}(\omega'),U^{u}(-\omega-\omega')^{*}]_{-} \:+ \:
U^{w}(\omega')[G^{u}(-\omega-\omega')^{*},U^{v}(\omega)]_{-}
\;\right) \:+\: (v,\omega\leftrightarrow w,\omega') \right\}
\end{aligned}\end{equation}
where $[\cdot,\cdot]_{-}$ denotes the commutator and $(v,\omega\leftrightarrow w,\omega')$ the simultaneous permutation of $v,w$ and $\omega,\omega'$ for the preceding terms within the trace. The XC term is defined as
\begin{equation}\label{betaDFT}\begin{aligned}
&\beta_{\text{xc}}^{uvw}(\omega,\omega')  = 
\sum_{\mu,\nu}\sum_{\sigma}P^{\sigma,u}_{\mu,\nu}(-\omega-\omega')^{*}V^{\sigma\:v,w}_{\nu,\mu}(\omega,\omega')-P^{\sigma,v,w}_{\mu,\nu}(\omega,\omega')V^{\sigma\:u}_{\nu,\mu}(-\omega-\omega')^{*} =
\\
&\sum_{\sigma,\sigma',\sigma''}\int n^{u}_{\sigma}(\bm{r},\omega+\omega')n^{v}_{\sigma'}(\bm{r}',\omega)n^{w}_{\sigma''}(\bm{r}'',\omega')
\left.\frac{\delta f_{\text{xc}}^{\sigma,\sigma'}(\bm{r},\bm{r}')}{\delta n_{\sigma''}(\bm{r}'')}\right\vert_{n^{(0)}_{\uparrow},n^{(0)}_{\downarrow}}d\bm{r}d\bm{r}'d\bm{r}''
\end{aligned}\end{equation}
where the second-order XC kernel in the last line is given for the GGA in \ref{AppC}. All quantities in \eqref{beta} are of first order, with $U^{u}(-\omega-\omega')^{*}=-U^{u}(\omega+\omega')^{T}$ and $G^{u}(-\omega-\omega')^{*}=G^{u}(\omega+\omega')^{T}$, hence the first-order solutions at $\omega$, $\omega'$ and $\omega+\omega'$ suffice to obtain the first hyperpolarizability $\beta^{uvw}(\omega,\omega')$. 

Expression \eqref{beta} is formally analogous to Refs. \cite{ADF2014,ADF2016}, 
and for $\eta=0$ reduces to the undamped version of Ref. \cite{Gisbergen1998}. This is also the structure of the quadratic response function in the CPP approach \cite{norman2005nonlinear}, in terms of first-order response vectors only. Other approaches to include the damping in the response functions can be found in the literature, but these generally differ in the non-linear regime from the present expressions using the relaxation mechanism. The adiabatic switching of the field is perhaps one of the most common ones, and is analyzed in \ref{AppA2}. Another route is to insert the damping a posteriori in the undamped response functions, then $\beta$ depends on the elimination rule used for the second-order response \cite{kristensen2009quasienergy}.

\textbf{Symmetry and subdomains of $\beta$.}
From \eqref{whermit} it follows that $\beta^{uvw}(-\omega,-\omega')=\beta^{uvw}(\omega,\omega')^{*}$.  
On the other hand, from \eqref{beta} it is clear that $\beta^{uwv}(\omega',\omega)=\beta^{uvw}(\omega,\omega')$, and one can also observe that the first two indices can be permuted in the event of
$\eta=0$, in which case $\beta^{vuw}(-\omega-\omega',\omega')=\beta^{uvw}(\omega,\omega')$. It may also be tempting to try relating $\beta^{vuw}(\omega,0)$ and $\beta^{uvw}(\omega,0)$, but the fact that generally $\Im U^{u}(0)\neq 0$ and $\Im G^{u}(0)\neq 0$ for $\eta\neq0$ precludes such equality. 

Although $\beta$ is a function of two independent frequencies, for the case of a monochromatic field with a steady component [namely $\omega_{l}\in\set{0,\omega_{1}}$ in \eqref{Efield}] only subdomains of $\beta$ with up to a single independent frequency [$\omega_{1}$] appear in the expansion \eqref{Cexpans} for the dipole \eqref{dipole}. There are 4 of those subdomains: 1) the static $\beta_{\text{STA}}^{uvw}=\beta^{uvw}(0,0)$; 2) the electro-optic or Pockels $\beta_{\text{POC}}^{uvw}(\omega)=\beta^{uvw}(\omega,0)$; 3) the rectification $\beta_{\text{REC}}^{uvw}(\omega)=\beta^{uvw}(\omega,-\omega)$; and 4) the second-harmonic hyperpolarizability $\beta_{\text{SHG}}^{uvw}(\omega)=\beta^{uvw}(\omega,\omega)$. From the above discussion, we observe that for $\eta=0$ it must be $\beta_{\text{REC}}^{uvw}(\omega)=\beta_{\text{POC}}^{vwu}(\omega)$, hence the electro-optic and rectification undamped effects would essentially be analogous. The relaxation mechanism is therefore what differentiates both contributions. It also follows that $\beta^{\parallel}_{\text{REC}}$ is always real-valued.

In analogy with the isotropic polarizability \eqref{alphaiso}, the first hyperpolarizability is often measured as a scalar quantity, its parallel component $\beta^{\parallel}$ along the direction of polarization \cite{Agren2005_comparison}. It can be expressed in terms of the tensor components in the original reference frame as
\begin{equation}\label{betaparal}
\beta^{\parallel}(\omega,\omega')=\frac{1}{5}\sum_{u,v}\frac{d^{u(0)}}{\abs{\bm{d}^{(0)}}}\left[
\beta^{uvv}(\omega,\omega') + \beta^{vuv}(\omega,\omega') + \beta^{vvu}(\omega,\omega')
\right]
\end{equation}
where $\bm{d}^{(0)}$ is the zeroth-order component of \eqref{dipole}, that is, the permanent dipole moment of the system. $\beta^{\parallel}$ is the projection on $\bm{d}^{(0)}$ of the vector part of the third-rank tensor $\beta^{uvw}$, and is therefore subject to stricter selection rules than the full $\beta^{uvw}$. The latter is strictly vanishing only for centrosymmetric point groups, whereas the former is vanishing for all point groups except the polar $C_{n}$, $C_{nv}$ ($n\geq1$). 

\subsection{Lanczos solver for the damped first-order equations}\label{secLanczos}
The self-consistent procedure of Section \ref{secCP} requires an individual resolution for every frequency, damping parameter and field direction. Here we describe an alternative procedure, the symmetric Lanczos method \cite{Brabec2015,Gruning2011_pseudoHermitianLanczos}, to solve the first-order TDGKS equations, and by extension obtain the first hyperpolarizabilities through the $(2n+1)-$rule, that requires an individual resolution only for each field direction. 
We show here how the Lanczos method can be applied as a solver for the first-order coupled-perturbed equations. In order to comply with the  standard notation in the Casida context (where Lanczos recursions are often applied), we introduce the excitation ($\mathrm{X}$) and de-excitation ($\mathrm{Y}$) amplitudes 
\begin{equation}\label{amplitudes}
\mathrm{X}_{ai}^{u}(\omega)=U_{a,i}^{u}(\omega) \:,\;\;
\mathrm{Y}_{ai}^{u}(\omega)=-U_{i,a}^{u}(\omega)
\end{equation}
which are treated as column vectors with $N_{\text{ov}}=N_{\text{occ}}N_{\text{virt}}$ entries, one for each field direction. Here the spin degree of freedom is embedded in the $a$ and $i$ states, which is a single index $\sigma$ since $a,i$ have equal spin in the collinear open-shell case [$N_{\text{ov}}=\sum_{\sigma}N^{\sigma}_{\text{occ}}N^{\sigma}_{\text{virt}}$]. 
Hereafter we denote vectors in this product space with upright bold font, e.g. $\mathbf{s}^{u}$ contains the elements $s^{u}_{a,i}$ in \eqref{sdip}. Square matrices acting on this space are denoted in blackboard bold, e.g. 
\begin{equation}\label{ABmat}\begin{aligned}
\mathbb{A}_{ai\sigma,bj\sigma'}&=(ai\vert bj)+(ai\vert f_{\text{xc}}^{\sigma,\sigma'}\vert bj)-\delta_{\sigma,\sigma'}(ab\vert ij)+\delta_{\sigma,\sigma'}\delta_{a,b}\delta_{i,j}[\varepsilon^{(0)}_{a}-\varepsilon^{(0)}_{i}]   \\
\mathbb{B}_{ai\sigma,bj\sigma'}&=(ai\vert bj)+(ai\vert f_{\text{xc}}^{\sigma,\sigma'}\vert bj)-\delta_{\sigma,\sigma'}(aj\vert bi)
\end{aligned}\end{equation}
where the bi-electronic integrals and the XC kernel (evaluated at the unperturbed spin densities) are expressed in the unperturbed MO basis, and the Fock-exchange is implicitly scaled or range-separated for the corresponding hybrid functionals. Note that $\mathbb{A}$, $\mathbb{B}$ are real symmetric matrices.

In terms of these, the first-order expression
\eqref{U1} can be written as the linear system of equations
\begin{equation}\label{ABsys}
\left[\begin{pmatrix}
\mathbb{A}&\mathbb{B}\\ \mathbb{B}&\mathbb{A}\end{pmatrix}-z\begin{pmatrix}\mathbb{I}&0\\ 0&-\mathbb{I}
\end{pmatrix}\right]
\begin{pmatrix}
\mathbf{X}^{u}\\ \mathbf{Y}^{u}\end{pmatrix}
=\begin{pmatrix}
\mathbf{s}^{u}\\ \mathbf{s}^{u}\end{pmatrix},\;\; z=\omega+i\eta,
\end{equation}
where we have employed \eqref{P1}, \eqref{H1}, \eqref{B1} and \eqref{V1}, noting that the off-diagonal blocks of $G^{u}(\omega)$ are linear in $\mathbf{X}^{u}$ and $\mathbf{Y}^{u}$. The self-consistent method outlined in Section \ref{secCP} is a fixed-point iteration on \eqref{ABsys}, for fixed $z$. These equations can also be solved straightforwardly for the amplitudes by diagonalization of a large $N_{\text{ov}}\times N_{\text{ov}}$ matrix, which
is the full-diagonalization Casida formulation of TDDFT \cite{Casida1995,CasidaHuixRotllant2012}, but the computational cost scales as $\mathcal{O}(N^{6})$ with system size and the memory requirements as $\mathcal{O}(N^{4})$. A common approximation is to neglect the coupling between excitations and de-excitations, $\mathbb{B}=0$, which decouples $\mathbf{X}^{u}$ and $\mathbf{Y}^{u}$. This is the Tamm-Dancoff approximation (TDA), not adopted in either of the two solvers of the present work.
The density \eqref{n1}, density matrix \eqref{P1} and polarizability \eqref{alpha} can be written in terms of the amplitudes as
\begin{equation}\label{PXY}
P^{u}_{\mu,\nu}(\omega)=\sum_{a,i}\left[c^{(0)}_{\mu,a}\mathrm{X}^{u}_{ai}(\omega)c^{(0)}_{\nu,i}+c^{(0)}_{\mu,i}\mathrm{Y}^{u}_{ai}(\omega)c^{(0)}_{\nu,a}\right]
\end{equation}
\begin{equation}\label{nXY}
n^{u}(\bm{r},\omega)=\sum_{a,i}\psi^{(0)}_{a}(\bm{r})\psi^{(0)}_{i}(\bm{r})\left[\mathrm{X}^{u}_{ai}(\omega)+\mathrm{Y}^{u}_{ai}(\omega)\right]
\end{equation}
\begin{equation}\label{alphaXY}
\alpha^{uv}(\omega)=(\mathbf{s}^{u})^{T}\left[\mathbf{X}^{v}(\omega)+\mathbf{Y}^{v}(\omega)\right]
\end{equation}
where $T$ denotes the transpose.

The system \eqref{ABsys} can be solved for the variables
$\mathbf{x}_{\pm}^{u}=\mathbf{X}^{u}\pm\mathbf{Y}^{u}$ as
\begin{equation}\label{xpm}
\mathbf{x}_{+}^{u}=2(\mathbb{A}-\mathbb{B})\left[\mathbb{L}-z^{2}\mathbb{I}\right]^{-1}\mathbf{s}^{u},  \;\;\;
\mathbf{x}_{-}^{u}=2z\left[\mathbb{L}-z^{2}\mathbb{I}\right]^{-1}\mathbf{s}^{u}, 
\end{equation}
\begin{equation}\label{Lmat}
\mathbb{L}=(\mathbb{A}+\mathbb{B})(\mathbb{A}-\mathbb{B})
\end{equation}
and therefore \eqref{alphaXY} can be written as
\begin{equation}\label{alphaLz}
\alpha^{uv}(\omega)=2(\mathbf{s}^{u})^{T}(\mathbb{A}-\mathbb{B})\left[\mathbb{L}-z^{2}\mathbb{I}\right]^{-1}\mathbf{s}^{v}
\end{equation}
If the matrix $\mathbb{M}=\mathbb{A}-\mathbb{B}$, whose entries are 
\begin{equation}\label{Mmat}
\mathbb{M}_{ai\sigma,bj\sigma'}=\delta_{\sigma,\sigma'}\{\delta_{a,b}\delta_{i,j}[\varepsilon^{(0)}_{a}-\varepsilon^{(0)}_{i}]+[(aj\vert bi)-(ab\vert ij)]\} 
\end{equation}
is positive definite \footnote{This is equivalent to the ground-state solution being stable against complex orbital rotations \cite{BauernschmittAhlrichs1996}. In the absence of Fock-exchange it is apparent that $\mathbb{M}$ is positive definite (and diagonal). With HF and hybrid functionals the condition held in all cases we tested.}, then $\langle\mathbf{x},\mathbf{y}\rangle_{\mathbb{M}}=\mathbf{x}^{T}\mathbb{M}\mathbf{y}$ defines an inner product with metric $\mathbb{M}$, and $\mathbb{L}$ is self-adjoint in it [$\langle\mathbf{x},\mathbb{L}\mathbf{y}\rangle_{\mathbb{M}}=\langle\mathbb{L}\mathbf{x},\mathbf{y}\rangle_{\mathbb{M}}$]. Its eigenvalues are then real and correspond to the squared excitation energies $\Omega_{I}^{2}$ of the Casida equation \cite{Casida1995} ($>0$ if $\mathbb{A}+\mathbb{B}$ is positive definite as well), and its eigenvectors $\bm{\zeta}_{I}$ can be chosen orthonormal in this inner product, $\langle\bm{\zeta}_{I},\bm{\zeta}_{J}\rangle_{\mathbb{M}}=\delta_{I,J}$, so that \eqref{alphaLz} can be expressed as 
\begin{equation}\label{spectral}
\alpha^{uv}(\omega)=2\sum_{I}\frac{\langle\mathbf{s}^{u},\bm{\zeta}_{I}\rangle_{\mathbb{M}}\langle\bm{\zeta}_{I},\mathbf{s}^{v}\rangle_{\mathbb{M}}}{\Omega_{I}^{2}-z^{2}}
\end{equation}
where frequency and damping enter only through $z^{2}=(\omega+i\eta)^{2}$. Iterative eigensolvers of Davidson type converge the lowest $\Omega_{I}$ and $\boldsymbol{\zeta}_{I}$ explicitly \cite{Davidson1975,Stratmann1998,Furche2016} and \eqref{spectral} is thus evaluated as a truncated sum. Instead, the Lanczos recursion approximates the whole sum at once, not directly targeting individual excitations.

\textbf{Lanczos recursion.}
Consider the Krylov subspace of order $k$ generated by $\mathbb{L}$ and $\mathbf{s}^{u}$, $\mathcal{K}_{k}^{u}=\text{Span}\{\mathbf{s}^{u},\mathbb{L}\mathbf{s}^{u},\dots,\mathbb{L}^{k-1}\mathbf{s}^{u}\}$. $\mathbb{L}$ \eqref{Lmat} is not a symmetric matrix but is self-adjoint with the inner product
$\langle\cdot,\cdot\rangle_{\mathbb{M}}$, which we adopt in the remainder of the section. A symmetric Lanczos algorithm \cite{GolubMeurant2010} can be formulated in this context
by building an orthonormal basis $\mathbf{q}_{1},\dots,\mathbf{q}_{k}$ [$\langle\mathbf{q}_{l},\mathbf{q}_{l'}\rangle_{\mathbb{M}}=\delta_{l,l'}$] of $\mathcal{K}_{k}^{u}$, constructed in such a way that $\mathbb{L}$ is represented by a tridiagonal matrix. The algorithm consists in a recursion over $k\geq1$, in each step adding a new vector to the orthonormal basis.
In particular, we formulate the three-term recurrence (or Lanczos chain) 
\begin{equation}\label{lanczos}
\mathbb{L}\mathbf{q}_{k}=b_{k}\mathbf{q}_{k+1}+a_{k}\mathbf{q}_{k}+b_{k-1}\mathbf{q}_{k-1}
\end{equation}
with $\mathbf{q}_{0}=0$ and the starting vector   $\mathbf{q}_{1}=\mathbf{s}^{u}/b_{0}$, $b_{0}=\norm{\mathbf{s}^{u}}_{\mathbb{M}}$ by construction of $\mathcal{K}_{1}^{u}$. The real-valued coefficients $a_{k}=\langle\mathbf{q}_{k},\mathbb{L}\mathbf{q}_{k}\rangle_{\mathbb{M}}$ and $b_{k}>0$ form the tridiagonal matrix
\begin{equation}\label{Tk}
\mathbb{T}_{k}=
\begin{pmatrix}
a_{1}&b_{1}&&0\\ b_{1}&a_{2}&\ddots&\\ &\ddots&\ddots&b_{k-1}\\ 0&&b_{k-1}&a_{k}
\end{pmatrix}
\end{equation}
which represents $\mathbb{L}$ in this basis,  $(\mathbb{T}_{k})_{l,l'}=\langle\mathbf{q}_{l},\mathbb{L}\mathbf{q}_{l'}\rangle_{\mathbb{M}}$. Storing $\mathbf{p}_{k}=\mathbb{M}\mathbf{q}_{k}$, the $k^{\text{th}}-$ iteration computes $\mathbf{t}_{k}=\mathbb{L}\mathbf{q}_{k}=(\mathbb{A}+\mathbb{B})\mathbf{p}_{k}$, from which $a_{k}=\mathbf{p}_{k}^{T}\mathbf{t}_{k}$ and the residual $\mathbf{r}_{k}=\mathbf{t}_{k}-a_{k}\mathbf{q}_{k}-b_{k-1}\mathbf{q}_{k-1}$ follow. From them, we form $\mathbb{M}\mathbf{r}_{k}$, $b_{k}=\norm{\mathbf{r}_{k}}_{\mathbb{M}}$, $\mathbf{q}_{k+1}=\mathbf{r}_{k}/b_{k}$ and $\mathbf{p}_{k+1}=\mathbb{M}\mathbf{r}_{k}/b_{k}$. That is, each iteration requires one matrix-vector product with $\mathbb{A}+\mathbb{B}$ and another with $\mathbb{M}$, which are computed without forming the $N_{\text{ov}}\times N_{\text{ov}}$ matrices, see Section \ref{secImpl}.

In this formulation, the diagonal elements of the polarizability \eqref{alphaLz} read 
\begin{equation}\label{alpha_innerp}
\alpha^{uu}(\omega)=
2b_{0}^{2}\langle\mathbf{q}_{1},[\mathbb{L}-z^{2}\mathbb{I}_{k}]^{-1}\mathbf{q}_{1}\rangle_{\mathbb{M}}
\approx 2b_{0}^{2}\left[(\mathbb{T}_{k}-z^{2}\mathbb{I}_{k})^{-1}\right]_{1,1}
\end{equation}
The last expression, which we denote as the approximant $\alpha^{uu}_{k}(\omega)$ becomes independent of $k$ at convergence. 
The entries of the inverse of a tridiagonal matrix can be expressed as continued fractions, in particular
\begin{equation}\label{cf}
\alpha^{uu}_{k}(\omega)
=\cfrac{2b_{0}^{2}}{a_{1}-z^{2}-\cfrac{b_{1}^{2}}{a_{2}-z^{2}-\cfrac{b_{2}^{2}}{\ddots-\cfrac{b_{k-1}^{2}}{a_{k}-z^{2}}}}}
\end{equation}
which provides a simple expression to approximate $\alpha^{uu}(\omega)$ at cost $\mathcal{O}(k)$ where frequency and damping appear only through $z^{2}$, hence the same set of Lanczos coefficients in the chain equation \eqref{lanczos} can be used to obtain $\alpha^{uu}(\omega)$ for any grid of $\omega$ and $\eta$. The underlying reason is that $\mathbb{L}-z^{2}\mathbb{I}$ generates the same Krylov subspace as $\mathbb{L}$. As a function of $z^{2}$, \eqref{cf} is a rational approximant with $k$ poles (at the eigenvalues of $\mathbb{T}_{k}$), which can be interpreted as a $k-$point Gauss quadrature of \eqref{spectral} \cite{Shao2018_SPLanczos}.
In practice, $k\ll N_{\text{ov}}$ (at $k=N_{\text{ov}}$ the approximation would be exact), and increasing the maximum $\omega$ or decreasing the minimum $\eta$ increases the number of iterations to convergence (see Section \ref{secResults}), nevertheless $k\sim10^{2}-10^{3}$ sufficed in all cases we tested.

\textbf{Amplitudes and first hyperpolarizability from the chain.}
The hyperpolarizability \eqref{beta} requires the first-order solutions $U^{u}(\omega)$ and the diagonal blocks of $G^{u}(\omega)$ at different frequencies [$0,\omega$ for $\beta_{\text{POC}}(\omega)$, $\beta_{\text{REC}}(\omega)$; $\omega,2\omega$ for $\beta_{\text{SHG}}(\omega)$; which constitute the frequency grid for $\beta$] for the selected damping. These can be reconstructed from a converged Lanczos chain as follows. From \eqref{xpm}, $\mathbf{X}^{u}=[\mathbb{M}+z\mathbb{I}]\boldsymbol{\xi}$, $\mathbf{Y}^{u}=[\mathbb{M}-z\mathbb{I}]\boldsymbol{\xi}$, with 
\begin{equation}\label{xiL}
\boldsymbol{\xi}=[\mathbb{L}-z^{2}\mathbb{I}]^{-1}\mathbf{s}^{u}
\end{equation}
and the Cartesian $u$ index is implicit. In the Lanczos basis, $\boldsymbol{\xi}$ is approximated by $\boldsymbol{\xi}\approx\sum_{l=1}^{k}\xi_{l}\mathbf{q}_{l}$, with the expansion coefficients $\xi_{l}=\langle\mathbf{q}_{l},\boldsymbol{\xi}\rangle_{\mathbb{M}}\approx b_{0}[(\mathbb{T}_{k}-z^{2}\mathbb{I}_{k})^{-1}]_{l,1}$. The amplitudes, and therefore $U^{u}(\omega)$, are then expressed as
\begin{equation}\label{xyrec}
\mathbf{X}^{u}(\omega)=\sum_{l=1}^{k}\xi_{l}(\mathbf{p}_{l}+z\mathbf{q}_{l}),\;\;
\mathbf{Y}^{u}(\omega)=\sum_{l=1}^{k}\xi_{l}(\mathbf{p}_{l}-z\mathbf{q}_{l})
\end{equation}
The coefficients $\xi_{l}$, which depend on $z$, can also be obtained exploiting the structure of the inverse of the tridiagonal matrix, at cost $\mathcal{O}(k)$ for each $\omega$ and $\eta$. In particular,
\begin{equation}\label{XY_coeffs}\begin{aligned}
&\xi_{1}=b_{0}\kappa_{1}, &&\xi_{j}=-b_{j-1}\kappa_{j}\xi_{j-1} \;\;(j=2,\dots,k), \\
&\kappa_{j}=\frac{1}{a_{j}-z^{2}-b_{j}^{2}\kappa_{j+1}}, &&\kappa_{k+1}=0
\end{aligned}\end{equation}
where $\kappa_{j}$, computed downwards in $j=k,\dots,1$, is the continued fraction of the form \eqref{cf} that starts at $a_{j}$, hence each $\xi_{j}$ is not evaluated as a continued fraction but as a product of them. Since $\kappa_{1}=[(\mathbb{T}_{k}-z^{2}\mathbb{I}_{k})^{-1}]_{1,1}$, $\alpha^{uu}_{k}(\omega)=2b_{0}\xi_{1}(z)$, which is precisely \eqref{alphaXY} evaluated with the amplitudes \eqref{xyrec}. The diagonal blocks of $G^{u}(\omega)$ then require one evaluation of $H^{u}(\omega)$ from $P^{u}(\omega)$, at the cost of one bi-electronic build per frequency and direction, after which the hyperpolarizability \eqref{beta} is evaluated exactly as for the self-consistent solver.

We can employ the spectrum $(\Omega_{I}^{2},\bm{\zeta}_{I})$ of $\mathbb{L}$ to obtain the pole structure of the first
hyperpolarizability, in analogy with the first-order expression \eqref{spectral}. Expanding $\boldsymbol{\xi}$ in the $\mathbb{M}$-orthonormal
basis $\set{\bm{\zeta}_{I}}$ and inserting it in the amplitudes $\mathbf{X}^{u}=[\mathbb{M}+z\mathbb{I}]\boldsymbol{\xi}$,
$\mathbf{Y}^{u}=[\mathbb{M}-z\mathbb{I}]\boldsymbol{\xi}$, we can write
\begin{equation}\label{XYspectral}
\begin{pmatrix}\mathbf{X}^{u}\\ \mathbf{Y}^{u}\end{pmatrix}=\sum_{I}\frac{\langle\mathbf{s}^{u},\bm{\zeta}_{I}\rangle_{\mathbb{M}}}{\sqrt{\Omega_{I}}}\left[
\frac{1}{\Omega_{I}-z}\begin{pmatrix}\mathbf{a}^{+}_{I}\\ \mathbf{a}^{-}_{I}\end{pmatrix}+
\frac{1}{\Omega_{I}+z}\begin{pmatrix}\mathbf{a}^{-}_{I}\\ \mathbf{a}^{+}_{I}\end{pmatrix}\right]
\end{equation}
where $\mathbf{a}^{\pm}_{I}=\frac{1}{2\sqrt{\Omega_{I}}}\left[\mathbb{M}\pm\Omega_{I}\mathbb{I}\right]\bm{\zeta}_{I}$. From \eqref{XYspectral} we observe that the amplitudes have poles at $z=\pm\Omega_{I}$, which translates into the diagonal blocks of $G^{u}$ since these are linear in the amplitudes (aside from the bare dipole term $-s^{u}$).
Every term of \eqref{beta}, including $\beta_{\text{xc}}$ [trilinear in the first-order densities \eqref{nXY}], is
therefore a product of at most one factor $1/(\Omega_{I}\pm z)$ per frequency argument, at
$z=\omega+\omega'+i\eta$ [through $U^{u}(-\omega-\omega')^{*}=-U^{u}(\omega+\omega')^{T}$], $z=\omega+i\eta$ and
$z=\omega'+i\eta$, with coefficients independent of $\omega$ and $\eta$. The resonances of $\beta$ then
occur at $\omega=\Omega_{I}$ for POC and REC, and at $\omega=\Omega_{I}$ and $2\omega=\Omega_{I}$ (two-photon) for SHG,
with the same damping as in $\alpha$ but generally different weights. Furthermore, $\beta$ decays with $\abs{\omega}$ and all poles of POC and SHG are in the lower complex $\omega$ plane, hence these two tensors satisfy the Kramers-Kronig relations \cite{BoydNLO} linking their real and imaginary parts, as the polarizability does. These relations do not hold for REC, since it has poles at both sides of the real axis, which is consistent with the fact that $\Im\beta^{\parallel}_{\text{REC}}=0$ unconditionally. 
 
A similar reconstruction has been used in the Liouville-Lanczos approach to give the perturbed density at selected frequencies \cite{Malcioglu2011_turboTDDFT}, though to our knowledge not the second-order response.

\section{Implementation} \label{secImpl}
The damped dynamical methods here described have been implemented in the \texttt{CRYSTAL} code \cite{crystal17,crystal23}, at present for finite systems. 
The work builds on previous implementations of the coupled-perturbed method for calculating the polarizability and hyperpolarizabilities both for static\cite{ferrero2008_1,ferrero2008_2,ferrero2008_3,orlando2010calculation,rerat2008comparison} and dynamic\cite{CRYSTALdyn1,CRYSTALdyn2,bernasconi2011first,rerat2016computation,rerat2018ab} dipolar electric fields, which were compatible with periodic systems but at zero damping \footnote{An exception being the approximate treatment of damping for the dynamic polarizability reported in Ref. \cite{CRYSTALdyn2}, which, however, we found to be inaccurate for calculating the optical absorption spectra of simple molecules, thus motivating this work.}.
Following the traditional approaches of the code, all bi-electronic contributions are computed with the direct method (i.e. non-resident in memory) in the AO basis, with a set of tolerances (\texttt{TOLINTEG} \cite{crystal23Manual}) to reduce their formal quartic scaling, while the first and second-order XC kernels are evaluated with the aid of the \texttt{XCFun} library \cite{Ekstrom2010_xcfun} (see \ref{AppC}) on the same integration grid of the ground-state calculation, based on the implementation of Ref. \cite{orlando2010calculation}. 

The bottleneck per iteration of both solvers is the evaluation of the bi-electronic term \eqref{B1}. As in the ground-state SCF, the integrals are contracted in the AO basis with the (now perturbed) density matrix and only the resulting matrices are transformed to the MO basis, hence 4-index transformations and the associated dense tensors are bypassed.
Since $P^{u}(\omega)$ is complex and not symmetric for $\eta\neq0$ \footnote{It is precisely this fact and the associated, generalized hermiticity relations \eqref{whermit} that complicate the adaptation of existing coupled-perturbed codes to the damped case, despite the formal similarities between the $\eta=0$ and $\eta\neq 0$ cases. This applies also to periodic codes like \texttt{CRYSTAL}, since the bi-electronic integrals in \eqref{B1} are contracted with the lattice components of $P^{u}(\omega)$, which in the absence of damping are real-valued along with the density in the XC term \eqref{V1}.}, the Fock-exchange part requires $P^{u}(\omega)$ and $P^{u}(-\omega)^{*}$ separately, whereas the Hartree and XC parts depend only on their sum through the density
\begin{equation}\label{dens_symmetrized}
n_{\sigma}^{u}(\bm{r},\omega)=\sum_{\mu\geq\nu}\frac{1}{1+\delta_{\mu,\nu}}[P^{\sigma\:u}_{\mu,\nu}(\omega)+P^{\sigma\:u}_{\mu,\nu}(-\omega)^{*}]\varphi_{\mu}(\bm{r})\varphi_{\nu}(\bm{r})
\end{equation}
In our implementation, this amounts to a $\sim30\%$ wall-time increase for hybrid functionals with respect to the unperturbed case (for global and range-separated hybrids alike), and virtually no extra cost for the Hartree and XC terms. The reduction of symmetry in $P^{u}(\omega)$ with respect to $P^{(0)}$ is largely compensated by exploiting the full unperturbed point group in the bi-electronic integrals, as explained in \ref{AppB} and based on the implementation in Ref. \cite{thesis_jacopo}. However, the multipolar treatment of the long-range Coulomb interaction and the bipolar expansion of the bi-electronic integrals have not yet been extended to the dynamical calculations, which in practice amounts to considering an infinite \texttt{TOLINTEG}(2) after the SCF. In this regard, we also observed that the usual setting \texttt{TOLINTEG}(4) $<$ \texttt{TOLINTEG}(5) for the truncation of the Fock-exchange integrals yielded small asymmetries in $\mathbb{A}$ and $\mathbb{B}$ that are amplified across Lanczos iterations. Hence, \texttt{TOLINTEG}(4) is automatically set to the requested \texttt{TOLINTEG}(5) in Lanczos calculations with hybrid functionals, unlike for the for the self-consistent solver (where this effect is harmless). 

\textbf{Self-consistent solver.} 
In order to reach self-consistency we employ a Pulay or direct inversion of the iterative subspace (DIIS) extrapolation \cite{Pulay1980} for the perturbed Hamiltonian $H^{u}(\omega)$. Both its real and imaginary parts are included on equal footing, and the error vector is given by the change of the occ-virt and virt-occ blocks of $G^{u}(\omega)$ between consecutive cycles. For the static (not damped) CPKS, this approach has been observed to increase the convergence speed significantly \cite{maschio2018DIIS,weber2003improved}, which we found to hold also for the present dynamical formulation (where a simple linear mixer failed to converge near resonances even in simple molecules, see the discussion on DIIS convergence in Section \ref{sec_hBN}). For the linear system \eqref{ABsys} at fixed $z$, DIIS is essentially equivalent to the GMRES Krylov-subspace method \cite{WalkerNi2011,RohwedderSchneider2011}, here preconditioned by the uncoupled denominators of \eqref{U1}. As for other solvers of the damped response equations \cite{Kauczor2011_CPP}, its convergence deteriorates beyond the absorption onset, where  $\omega$ is generally close to many excitation energies and the uncoupled preconditioner is less effective. Therefore, the number of cycles tends to increase with $\omega$, which makes high-harmonic hyperpolarizabilities more computationally demanding than other optical tensors (at equal frequency); and also with system size, to the point that even solving for a single $\omega$ well past the absorption onset becomes impractical, as demonstrated below. 

\textbf{Lanczos solver.}
The bi-electronic and XC builds for the matrix-vector products of each step are computed with the same routines as for the self-consistent solver, so that $\mathbb{A}$ and $\mathbb{B}$ are never formed.
A real-valued vector $\mathbf{v}$ in the product space, in particular $\mathbf{q}_{k}$, $\mathbf{p}_{k}$ or $\mathbf{r}_{k}$, defines the AO matrix $D^{\sigma}_{\mu,\nu}[\mathbf{v}]=\sum_{a,i}c^{\sigma\:(0)}_{\mu,a}\mathrm{v}^{\sigma}_{ai}c^{\sigma\:(0)}_{\nu,i}$, hence
\begin{equation}\label{ApB}\begin{aligned}
[(\mathbb{A}+\mathbb{B})\mathbf{v}]_{ai\sigma}=&(\varepsilon^{\sigma (0)}_{a}-\varepsilon^{\sigma (0)}_{i})\mathrm{v}^{\sigma}_{ai}+
\sum_{\rho,\tau}c^{\sigma\:(0)}_{\rho,a}c^{\sigma\:(0)}_{\tau,i}
\sum_{\mu,\nu,\sigma'}\left\{D^{\sigma'}_{\mu,\nu}[\mathbf{v}]+D^{\sigma'}_{\nu,\mu}[\mathbf{v}]\right\}\cdot
\\
&\cdot[(\rho\tau\vert\mu\nu) -\delta_{\sigma,\sigma'}(\rho\mu\vert \tau\nu)
+(\rho\tau\vert f^{\sigma\sigma'}_{\text{xc}}\vert\mu\nu)] 
\end{aligned}\end{equation}
\begin{equation}\label{AmB}\begin{aligned}
[(\mathbb{A}-\mathbb{B})\mathbf{v}]_{ai\sigma}=(\varepsilon^{\sigma (0)}_{a}-\varepsilon^{\sigma (0)}_{i})\mathrm{v}^{\sigma}_{ai} - \sum_{\rho,\tau}c^{\sigma\:(0)}_{\rho,a}c^{\sigma\:(0)}_{\tau,i}
\sum_{\mu,\nu}\left\{D^{\sigma}_{\mu,\nu}[\mathbf{v}]-D^{\sigma}_{\nu,\mu}[\mathbf{v}]\right\}(\rho\mu\vert \tau\nu)
\end{aligned}\end{equation}
where the contraction with the MO coefficients to yield the $(a,i)$ indices is performed after the contraction of the AO integrals with $D$. Note that these expressions contain the bi-electronic term \eqref{B1} evaluated with $D^{\sigma'}_{\mu,\nu}[\mathbf{v}]\pm D^{\sigma'}_{\nu,\mu}[\mathbf{v}]$ in place of the density matrix, and likewise with the XC term \eqref{V1}, where the density matrix is replaced within the density. The anti-symmetric combination vanishes upon contraction with the Hartree and XC terms, leaving only the Fock-exchange in \eqref{AmB}. Therefore, a Lanczos iteration involves two bi-electronic builds with HF/hybrid-functionals and just one with pure semi-local functionals. In contrast, the self-consistent solver involves a single build per cycle regardless of the functional. A single XC build is required per iteration in both methods. 

It is well known that the mutual orthogonality of the basis vectors $\mathbf{q}_{l}$, $l=1,\dots,k$ may degrade after a number of iterations due to finite precision arithmetic \cite{MeurantStrakos2006,tretiak2009representation}, which may make long chains numerically unstable. %and also yield spurious copies of poles. 
In order to avoid this issue we explicitly re-orthogonalize every new $\mathbf{q}_{k}$ against $\mathbf{q}_{k-1},\dots,\mathbf{q}_{1}$ after a certain number of iterations (for the calculations in this work, in every iteration), at negligible cost compared with the bi-electronic builds. Note that $\mathbf{q}_{l}$, $l=1,\dots,k$, are in any case needed for the reconstruction of the amplitudes \eqref{xyrec}, alongside $\mathbf{p}_{l}$, $l=1,\dots,k$. 
On the other hand, we provide the option to restrict the product space from the full $N_{\text{ov}}$ to the occ-virt pairs with $\varepsilon^{(0)}_{a}-\varepsilon^{(0)}_{i}$ below a cutoff, zeroing the corresponding entries of the matrices and vectors. This procedure targets the removal of excitations from core orbitals and excitations into high-lying virtual orbitals, whose contribution to $\alpha$ in the optical range tends to be small but require a larger number of iterations for convergence \cite{Brabec2015}. 

The convergence of $\alpha$ is evaluated at checkpoints every $m$ iterations, through the largest relative change of $\alpha^{uu}_{k}$ (over the selected grid of $\omega$ and $\eta$) measured from the previous checkpoint. For $\beta$, in addition, we require that the reconstructed amplitudes \eqref{xyrec} also converge, which we quantify directly with the residual $b_{k}\xi_{k}\mathbf{q}_{k+1}$, obtained 
by applying \eqref{lanczos} in \eqref{xiL} and identifying the contribution from the newest iteration. This quantity is evaluated in each checkpoint for every frequency of the $\beta$ grid, in second-order calculations. The state of the Lanczos chain is saved at every checkpoint to enable the restart in subsequent jobs, namely by writing $\mathbf{q}_{l},\mathbf{p}_{l},a_{l},b_{l}$, $l=1,\dots,k$ per direction, as well as the MO coefficients (to guarantee that the gauge is fixed), to disk.

\section{Results and discussion}\label{secResults}

\subsection{Validation in a small molecule}\label{sec_comparison}

\begin{figure}[H]
\centering 
\includegraphics[width=1.0\textwidth]{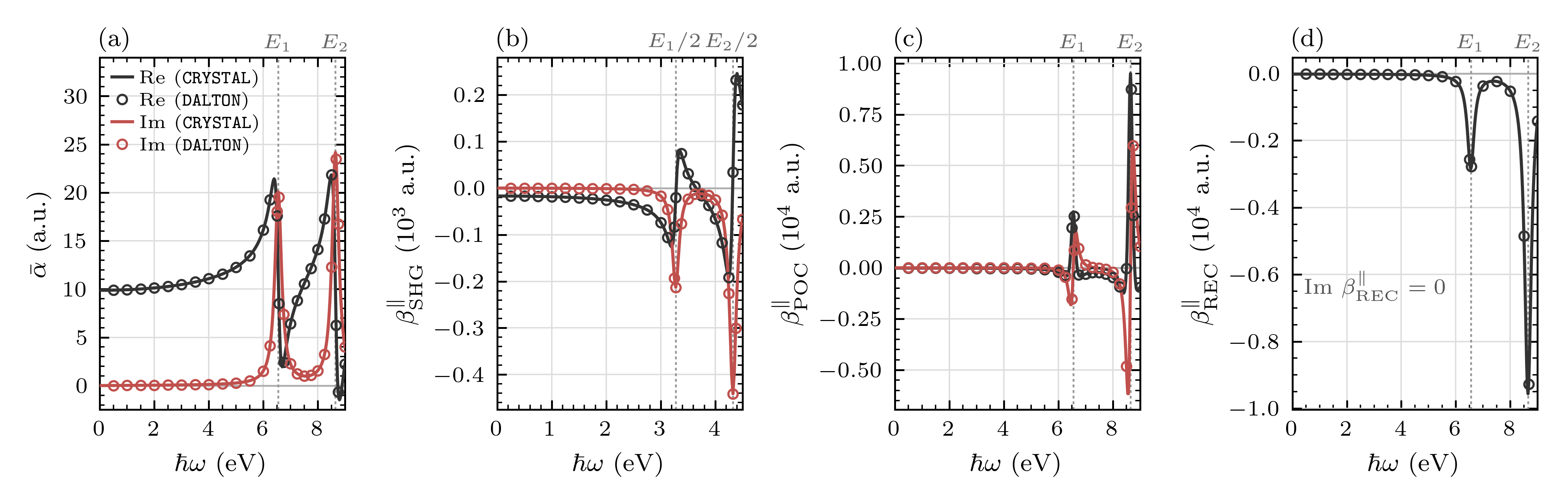}
\caption{\label{FigH2O} Polarizability and first-hyperpolarizability of H$_2$O (SVWN LDA functional, aug-cc-pVDZ basis, $\eta=0.15$ eV damping) from both the present implementation in \texttt{CRYSTAL} (Lanczos solver, solid lines)
and the implementation in \texttt{DALTON} (open circles).
(a) Isotropic polarizability $\bar{\alpha}(\omega)$, \eqref{alphaiso}. (b)$-$(d) First
hyperpolarizability $\beta^{\parallel}$ along the dipole moment direction ($z$), \eqref{betaparal}, at the second-harmonic generation (SHG), Pockels (POC) and rectification (REC) subdomains. 
Dotted lines: energies of the lowest bright excitations, $E_{1}=6.55$ eV  and $E_{2}=8.65$ eV; in (b) they are placed at half the values, corresponding to the two-photon resonances of SHG. The maximum relative deviation between both codes is of the order $10^{-5}$ at all plotted points.}
\end{figure}

Our coupled-perturbed implementation was tested against the molecular code \texttt{DALTON} (Dalton2020.1 release), which provides the polarizability and first hyperpolarizability with an iterative eigensolver \cite{Aidas2014_Dalton,norman2005nonlinear}, a different route to the same physical quantities. As a canonical example we consider the
water molecule (whose point group $C_{2v}$ allows for a permanent dipole and thus a finite $\beta^{\parallel}$), at the LDA level with the SVWN functional (VWN5 correlation \cite{VMNcorr}), employing the
aug-cc-pVDZ basis set \cite{cc_pVTZ,kendall1992electron,pritchard2019new} (whose diffuse functions are required for a proper description of $\beta$ in small molecules \cite{Agren2002_beta}) and damping factor $\eta=0.15$ eV. The ground-state total energies differ by
$10^{-6}$ Ha. In our implementation we employed the Lanczos solver in the full product space of $N_{\text{ov}}=180$ (no energy cutoff), with conservatively tight convergence criteria of $10^{-6}$ maximum relative change of $\alpha^{uu}$ for all frequencies between consecutive checkpoints (every 10 iterations), and likewise for the residual of the reconstructed
amplitudes $\mathbf{X},\mathbf{Y}$ required for $\beta$. 
The 3 chains (one per Cartesian direction), then converged in 90$-$110 iterations, each providing the response on a grid of 360
frequencies, whereas in \texttt{DALTON} the response equations were solved separately for each sampled frequency.  

The peaks are very well-defined in such a small molecule, as each arises from an individual excited state, with energy gaps much larger than $\eta$. The two lowest bright excitations are $E_{1}=6.55$ eV, polarized along $y$ (perpendicular to the molecular plane), and $E_{2}=8.65$ eV, polarized along $z$ (parallel to the dipole moment). 
The results from both codes, shown in Figure \ref{FigH2O}, are in agreement up to a maximum relative deviation of $3\cdot 10^{-5}$, among all  components $\abs{\alpha^{uv}}$ and $\abs{\beta^{uvw}}$ and all frequencies. This includes the resonant regions, comprised of the one-photon resonances of $\beta_{\text{POC}}$, $\beta_{\text{REC}}$ at $E_{1}$ and $E_{2}$, and the two-photon resonances of $\beta_{\text{SHG}}$ at $E_{1}/2$ and $E_{2}/2$. In this case, the two excited states are bright in both the first and second-order responses, and $\alpha$, $\beta_{\text{SHG}}$, $\beta_{\text{POC}}$ display the usual shapes given by the Kramers-Kronig relations at isolated resonances.
Moreover, both codes yield a vanishing $\Im\beta^{\parallel}_{\text{REC}}$ (as it should be, regardless of symmetry), and the three $\beta$ subdomains converge to the
same static value $\beta^{\parallel}(0)=-17.0$ (a.u.). Likewise, in the static limit at the linear order we recover the value $\bar{\alpha}(0)=9.87$ (a.u.).

\subsection{Application: $\alpha$ and $\beta$ in hBN flakes}\label{sec_hBN}

\begin{figure}[H]
\centering 
\includegraphics[width=1.0\textwidth]{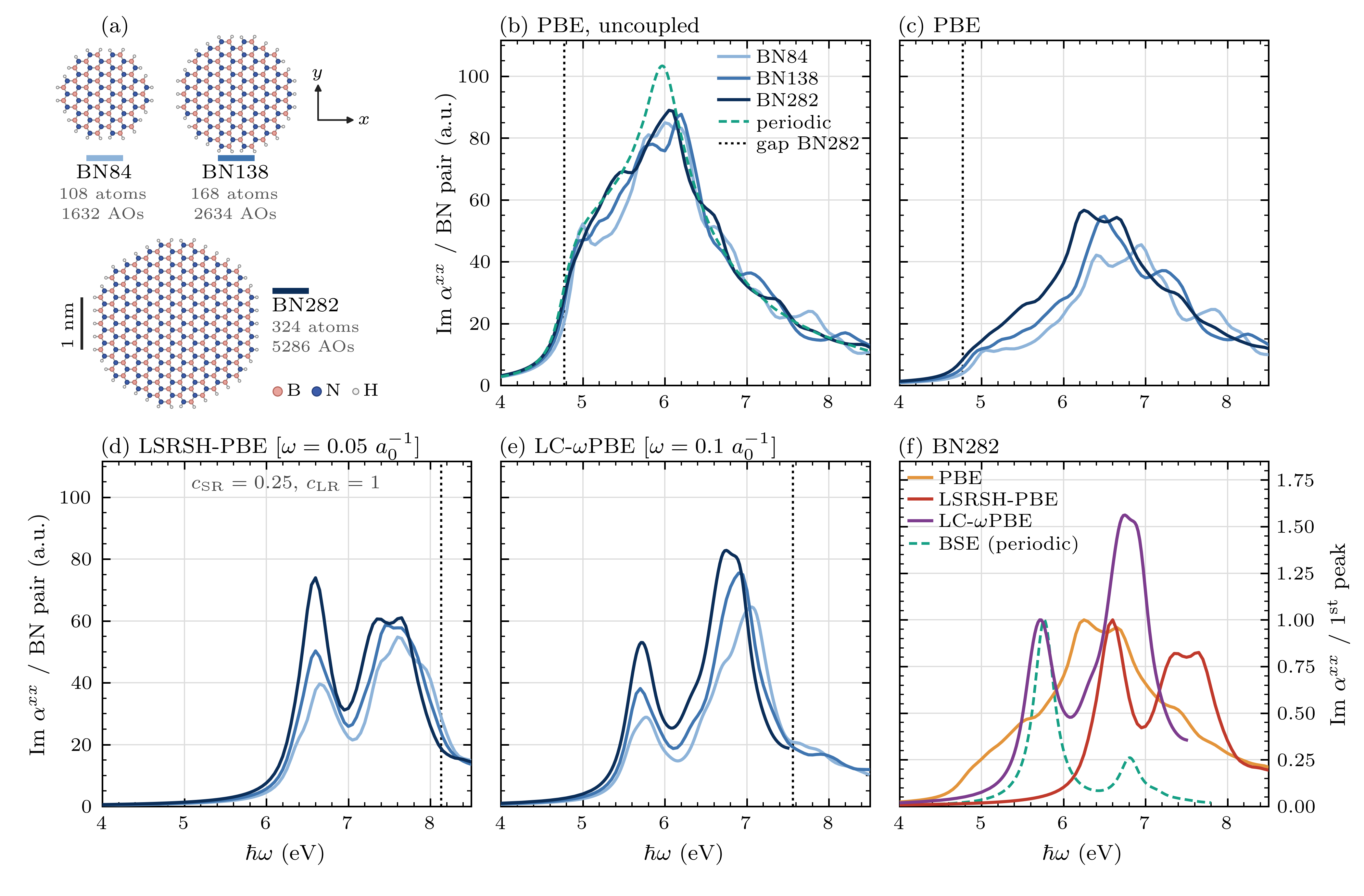}
\caption{\label{FigAlpha} Linear in-plane absorption for a series of 3 hBN flakes (pob-TZVP-rev2 basis, $\eta=0.15$ eV damping).
(a) The clusters BN84, BN138 and BN282, labeled by the total number of B, N atoms and depicted at scale. 
(b)$-$(e) $\Im\alpha^{xx}$ per B$-$N pair, in atomic units and with the color labels indicated in (a):
(b) uncoupled PBE (independent particle approximation), including the periodic monolayer (dashed);
(c) PBE; (d) LSRSH-PBE, with range-separation parameter $\omega=0.05$ $a_0^{-1}$ and Fock-exchange fractions
$c_{\mathrm{SR}}=0.25$ (short range) and $c_{\mathrm{LR}}=1$ (long range); (e) LC-$\omega$PBE, with range-separation parameter $\omega=0.1$ $a_0^{-1}$. Dotted lines in (b)$-$(e): HOMO-LUMO gap in the largest flake.
(f) Coupled spectra for BN282 alongside the BSE spectrum of the monolayer (dashed), each normalized to its first
maximum.
All coupled spectra were obtained with the Lanczos solver. 
Periodic uncoupled and BSE spectra computed with \texttt{XatuGTF} using the pob-TZVP-rev2 basis, the latter employing the TDA and a scissor correction over the PBE ground-state.}
\end{figure}

\textbf{hBN flakes.} As an application of the implementation we study the evolution of the polarizability and first hyperpolarizability in a series of 3 relatively large hexagonal boron nitride (hBN) planar flakes. The hBN clusters are depicted in Figure \ref{FigAlpha}a, and labeled as BN84, BN138 and BN282 according to the total number of B and N atoms they contain. They were constructed from the periodic hBN monolayer by keeping all atoms within a given radius from an hexagon center, restricting the radii to those for which every boundary atom retains exactly two first neighbors. This way, each of the edge B and N atoms is passivated with a single H atom. The series of flakes was selected to have the same edge topology, in which only the number of isolated edge sites (defined as 2-coordinate atoms whose two nearest neighbours are both 3-coordinate) grows with flake size (0, 6, 18 for BN84, BN138, BN282 respectively), while the remaining 24 edge atoms are identical in number and local environment across the series.
The B$-$N bond length, 1.449 \AA, was taken from the optimized periodic geometry, while the position of the H atoms was relaxed in each cluster, all geometry optimizations carried out at the PBE level \cite{Perdew1996_PBE} with otherwise the same SCF production settings as in the response calculations. This construction preserves the $D_{3h}$ point group symmetry of the periodic 2D structure, which forces $\alpha^{xx}=\alpha^{yy}$ (the in-plane response) and $\beta^{xxx}=-\beta^{xyy}=-\beta^{yxy}=-\beta^{yyx}$, where $x$ contains a twofold rotation symmetry as depicted in Figure \ref{FigAlpha}a \cite{garcia2023shift}. We assume that the clusters are fixed in space (like a solid), hence we consider $\alpha^{xx}$ and $\beta^{xxx}$ instead of the angle-averaged $\overline{\alpha}$ and $\beta^{\parallel}$ (which, as the permanent dipole, would vanish).

\textbf{Calculation settings.}
For the hBN systems, including all flake sizes and the periodic monolayer, we employed the basis set pob-TZVP-rev2 \cite{vilela2019bsse}, in the flakes removing the most diffuse $s$ function for the H atoms in order to reduce the sampling of the vacuum in the plane (whose impact is meant to be mitigated by the hydrogenation itself). The use of this basis set, designed for periodic systems, in the flakes allows to circumvent the linear dependencies introduced by other medium-size molecular basis sets such as def2-TZVP or cc-pVTZ. For the flakes we employed the truncation parameters \texttt{TOLINTEG} 10 10 10 200 200 for the bi-electronic integrals (the second playing no role in the dynamical coupled-perturbed calculations, as explained in Section \ref{secImpl}), which yield a relative convergence of $\alpha^{xx}$ no worse than $0.02\%$, as measured in BN282 for $\omega=5.3$ eV and $\eta=0.15$ eV. We employ this damping factor for the production calculations, being a good compromise between peak resolution and convergence speed. 
The Lanczos chains were checked every 25 steps and stopped when the largest relative change of $\alpha^{uu}$ among the frequency grid was below $10^{-3}$. For the $\beta$ calculations, in addition, we imposed that the relative residual of the reconstructed amplitudes $\mathbf{X},\mathbf{Y}$ be below $10^{-2}$ at $\omega=0$ and every $\omega$ of the grid. SHG values are computed up to half the maximum frequency, since convergence is required up to $2\omega$ for the response at $\omega$. The Lanczos product space of occ-virt pairs was truncated by the energy cutoff  $\varepsilon^{(0)}_{a}-\varepsilon^{(0)}_{i}<100$ eV. This truncation lowered $\Re\alpha$ by up to $0.2\%$ and changed $\Im\alpha$ by less than $0.1\%$ across the frequency range, with respect to the self-consistent solver, in the BN84 flake with PBE and $\eta=0.2$ eV.

\textbf{First-order response.}
The uncoupled response, also known as sum-over-states (SOS) or independent-particle approximation (IPA), is obtained by fixing $G^{u}=-s^{u}$ for the off-diagonal blocks, namely neglecting the field-induced variation of the density. The response tensors are then obtained in one-shot from cycle 0 without any self-consistency. This is done in Figure \ref{FigAlpha}b for the PBE functional, which shows that the uncoupled absorption is relatively independent of the flake size beyond the chosen minimum (BN84, where some oscillations due to state discretizations are still present). In fact, the agreement between the largest flake and the infinite size limit (2D lattice) is rather good, with a shift below 50 meV between the main peaks. In the coupled-perturbed calculation with the same functional, Figure \ref{FigAlpha}c, the global maxima are blue-shifted and weakened, and the effect of the system size is somewhat more significant. 
Nevertheless, the coupled-perturbed PBE spectra qualitatively resemble their uncoupled counterparts, with a single broad structure (to which many excitations contribute, forming a quasi-continuum) starting at the HOMO$-$LUMO gap, lacking a distinct dominant peak. 

However, it is known in the computational literature that the absorption of hBN in the monolayer phase is largely dominated by a first peak around 6 eV, arising from a strongly-bound exciton with binding energy $\sim 2$ eV \footnote{In the terminology of this work, this is just an excited state whose energy is measured backwards from the HOMO-LUMO gap.}, followed by a separated secondary peak about $\sim 1$ eV higher, to which two  bound excitons contribute \cite{zhang2022intervalley,galvani2016excitons}.
This picture is provided by the Bethe-Salpeter equation (BSE), where the static screened-Coulomb interaction plays the role of the Fock-exchange plus the adiabatic semi-local XC kernel \footnote{The Hartree term is also present in the BSE in the exact same form, where it is referred to as the exchange term. It is included in the singlet solutions, which are optically active.}. 
It is commonly regarded as a reference method, particularly for periodic systems \cite{Onida2002,MartinIE}, but it is considerably more expensive in general.
With a semi-local functional the coupling in \eqref{ABmat} depends only on the perturbed density, through the Hartree and XC terms, whereas the Fock-exchange depends on the full perturbed density matrix, which is a non-local behavior that more closely resembles that of the screened Coulomb interaction. Moreover, the diagonal of the Fock-exchange matrix in \eqref{ABmat} is $-(aa\vert ii)$, the Coulomb attraction between the (excited) electron at $a$ and the hole at $i$. 
A global hybrid includes a fixed fraction of Fock exchange (e.g. $25\%$ for PBE0 \cite{Adamo1999_PBE0}) at all distances, which amounts to an electron$-$hole attraction uniformly reduced with respect to the bare Coulomb interaction. In a 2D layer surrounded by vacuum, however, the screening is highly non-local (the effective interaction in the layer is reduced at short distances), but it recovers the unscreened Coulomb potential at long range \cite{Cudazzo2011}. 
Guided by these considerations, we test two range-separated hybrid functionals: LC-$\omega$PBE \cite{weintraub2009long,Vydrov2006_LCwPBE}, which replaces the PBE exchange by the Fock-exchange at long range (with the crossover set by the range-separation parameter $\omega$), and LSRSH-PBE \cite{crystal23Manual}, which employs the fractions $c_{\text{SR}}$ and $c_{\text{LR}}$ of Fock-exchange at short and long range, and reduces to LC-$\omega$PBE for $c_{\text{SR}}=0$, $c_{\text{LR}}=1$ and the given $\omega$.
For LC-$\omega$PBE we chose $\omega=0.1$ $a_{0}^{-1}$ to reproduce the first peak of our periodic BSE reference at 5.76 eV (see below), while for LSRSH-PBE we set $\omega=0.05$ $a_{0}^{-1}$, $c_{\text{SR}}=0.25$, $c_{\text{LR}}=1$ (full Fock-exchange at long range) so that the HOMO$-$LUMO gap of BN282 lies close to the GW gap of the monolayer (8.03 eV at the $K$ point \cite{zhang2022intervalley}). We keep the LC-$\omega$PBE and LSRSH-PBE parametrizations fixed across the flake series. Table \ref{TableE_hBN} summarizes the calculated data in the flake series for different functionals, including PBE0 (not displayed in the figures since we did not converge it fully in BN282, just enough to stabilize the first peak position). 

\begin{table}
\renewcommand{\arraystretch}{1.5}
\caption{\label{TableE_hBN} HOMO$-$LUMO gap $E_{g}$, first peak position $E_{1}$ in $\Im\alpha^{xx}$ and binding energy $E_{g}-E_{1}$ for the hBN flakes, for $\eta=0.15$ eV. $E_{1}$ is omitted for PBE due to its different absorption profile.}
\label{tabPeaks}
\begin{tabular}{|l|ccc|ccc|ccc|}
\hline
 & \multicolumn{3}{c|}{BN84} & \multicolumn{3}{c|}{BN138} & \multicolumn{3}{c|}{BN282} \\
\hline
 & $E_{g}$ & $E_{1}$ & $E_{g}-E_{1}$ & $E_{g}$ & $E_{1}$ & $E_{g}-E_{1}$ & $E_{g}$ & $E_{1}$ & $E_{g}-E_{1}$ \\
\hline
PBE & 4.97 & - & - & 4.89 & - & - & 4.77 & - & - \\
PBE0 & 7.22 & 6.59 & 0.63 & 7.13 & 6.52 & 0.61 & 7.01 & 6.50 & 0.51 \\
LSRSH-PBE & 8.34 & 6.66 & 1.68 & 8.25 & 6.60 & 1.65 & 8.13 & 6.59 & 1.54 \\
LC-$\omega$PBE & 7.77 & 5.77 & 2.0 & 7.68 & 5.70 & 1.98 & 7.56 & 5.72 & 1.84 \\
\hline
\end{tabular}
\end{table}

The coupled perturbed spectra with the range-separated hybrids, Figures \ref{FigAlpha}d and \ref{FigAlpha}e, indeed display 2-3 well separated peaks, the second and third close-lying and at similar amplitudes. The first peak now lies below the HOMO$-$LUMO gap in both cases ($E_{g}-E_{1}=1.84$ eV for LC-$\omega$PBE and 1.54 eV for LSRSH-PBE), which illustrates the impact of the Fock-exchange in contrast with the semi-local PBE response. This is also the case for PBE0, see Table \ref{TableE_hBN}, but the effect is weaker due to its smaller fraction of long range Fock-exchange. Unlike in the periodic case, there is many-to-one instead of a one-to-one correspondence between excited states and these absorption peaks, e.g. five excitations between 5.57 and 5.80 eV contribute to the first peak in BN282 with LC-$\omega$PBE. As a reference, we compare in Figure \ref{FigAlpha}f the absorption in BN282 for the different functionals with the periodic BSE result in the monolayer. The latter was obtained with \texttt{XatuGTF} \cite{garcia2025first,UriaAlvarez2024_Xatu}, employing the TDA for the singlet configuration and the same Gaussian basis set as for the flakes, along with the auxiliary basis def2-TZVP-RIFIT \cite{weigend1998ri} for the resolution of the identity  approximation, a basis for the electron-hole pairs made of $72\times72$ $\bm{k}-$points, two valence and one conduction band, and a ground-state reference given by the PBE calculation plus a (rigid) scissor correction of 3.27 eV to match the GW gap (as above with the LSRSH-PBE parametrization). The binding energy in this case comes out too large, at 2.27 eV, which is essentially an effect of the SCF and auxiliary basis sets and to a lesser extent of the $\bm{k}-$grid (it was converged to 1.88 eV in Ref. \cite{garcia2025first}), nevertheless it aims to provide a reference in the most similar conditions to the flakes. It can be observed that the position of the two peaks  is well reproduced by the LC-$\omega$PBE choice, but the relative amplitude is certainly not, at 1.56 vs 0.26 of BSE for the $2^{\text{nd}}/1^{\text{st}}$ peaks. In contrast, the LSRSH-PBE result fails to match the peak positions (starting at a similar HOMO$-$LUMO gap as the BSE$+$scissor calculation), although the peak separation is similar and the relative amplitude is better, 0.83, though still far from the periodic BSE result. 
The trade-off of LSRSH-PBE with respect to LC-$\omega$PBE comes from the short-range fraction of Fock-exchange, acting mainly within a radius $1/\omega$ (here covering several B$-$N bonds), which is a different approximation to the screening potential than the semi-local one but not always better. In these systems, it improves one aspect (amplitude ratio, namely the oscillator strengths) but worsens another (peak positions / binding energies). 
For the periodic hBN monolayer, a blue-shift of $\sim0.6$ eV in the first peak 
was found in the literature \cite{Ramasubramaniam2019_SRSH,Camarasa2023_SRSH} between 
TDDFT with a tuned screened range-separated hybrid (fitted to the GW@LDA gaps of hBN monolayer and bulk) and GW-BSE. 

Besides the difference in physical methods, the other important difference of the flakes with the periodic system is the finite size. As observed from the results across the series, $E_{1}$ varies by less than 0.1 eV and $E_{g}$ by about 0.2 eV, which is close to the periodic band gap (e.g. $E_{g}=7.50$ eV with LC-$\omega$PBE). In contrast, the amplitude ratio still changes significantly: for LC-$\omega$PBE 2.23 $\to$ 1.98 $\to$ 1.56, for LSRSH-PBE 1.38 $\to$ 1.17 $\to$ 0.83, sorted by increasing size. 

\begin{figure}[H]
\centering 
\includegraphics[width=1.0\textwidth]{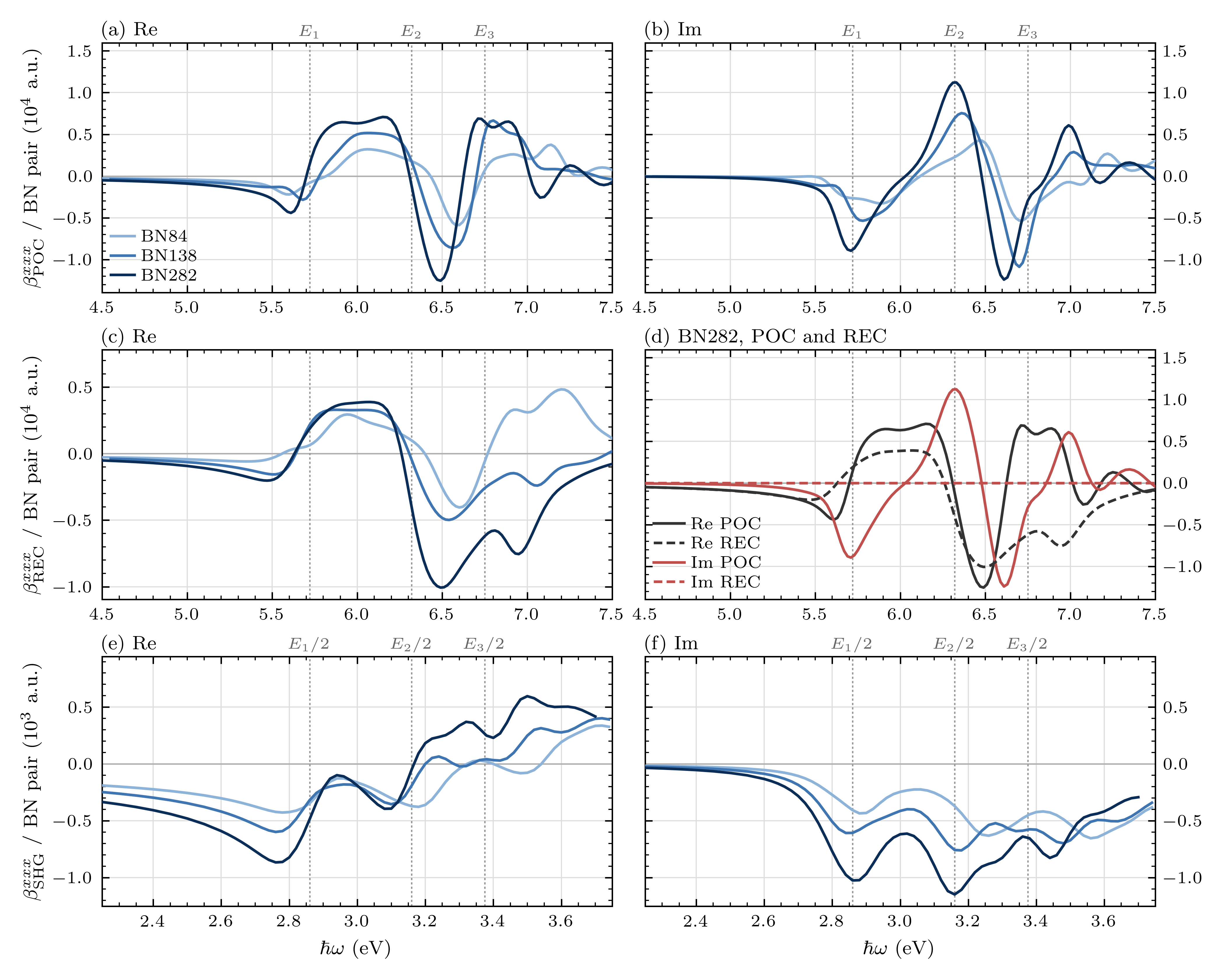}
\caption{\label{FigBeta} First hyperpolarizability frequency subdomains (per B$-$N pair) in the three hBN flakes (LC-$\omega$PBE with range-separation $\omega=0.1$ $a_0^{-1}$, pob-TZVP-rev2 basis, $\eta=0.15$ eV damping). The depicted component is the only independent entry of the $\beta$ tensor under the $D_{3h}$ point group, fulfilling $\beta^{xxx}=-\beta^{xyy}=-\beta^{yxy}=-\beta^{yyx}$.
(a), (b) Pockels (POC) real and imaginary parts; (c) rectification (REC) real part [the imaginary part is vanishing]; (d) POC and REC comparison, whose difference is purely due to the damping; (e), (f) second harmonic generation (SHG) real and imaginary parts.
Dotted lines: the absorption maxima of BN282, including the peaks at $E_{1}=5.72$ eV and $E_{3}=6.75$ eV, and the shoulder at $E_{2}=6.32$ eV.
The SHG frequency window is half that of POC and REC, in order to line up the two-photon resonances with the one-photon ones. $\beta$ obtained through the $(2n+1)-$rule from the first-order solution provided by the Lanczos method.
}
\end{figure}

\textbf{Second-order response.}
Next we consider the first hyperpolarizability in the flake series. Since the positions of the absorption peaks are directly relevant for the resonances of $\beta$, we employ the previous LC-$\omega$PBE parametrization, which means that the spectra past the first resonance at $E_{1}$ should not be compared to a periodic many-body result, but they are nevertheless insightful and valid at the (non-TDA) adiabatic TDDFT level for the finite flakes. As discussed below \eqref{XYspectral}, the resonances of $\beta$ are those of $\alpha$, appearing also at half the frequencies (two-photon) for SHG, but generally with a different amplitude. This is observed in the imaginary parts of Figure \ref{FigBeta}, which present either well defined minima or maxima, or a shoulder, at $E_{1}=5.72$ eV, $E_{2}=6.32$ eV (a shoulder in $\Im\alpha^{xx}$) and $E_{3}=6.75$ eV for $\Im$ POC (Figure \ref{FigBeta}b) and at $E_{1}/2$, $E_{2}/2$ and $E_{3}/2$ for $\Im$ SHG (Figure \ref{FigBeta}f). We note that the alignment of these features with the indicated $E_{1},E_{2},E_{3}$ need not be exact, since these represent the position of the peaks and in our large finite systems they are formed by multiple, close-lying excited states (nevertheless we loosely refer to these $E_{i}$ as excitation energies for simplicity). 
It is indeed apparent that the contribution of each excited state to the first and second-order response is very different, for instance, $E_{2}$ yields a weak shoulder in the absorption but a strong maximum in $\Im$ POC, whereas $E_{3}$ has the opposite behavior. This is not uncommon in the nonlinear response, where the contribution of each state is a product of three oscillator strengths. %(including one between virtual states). 
For instance, in the shift current (essentially the periodic analogue of REC) of 2D monolayers such couplings can make excitations that are dark in the absorption contribute to the nonlinear response, with a strength comparable to  that those that contribute most in the linear order \cite{Esteve2025_shift}.

The SHG results span the frequencies of the two-photon but not the one-photon resonances, which require convergence at frequencies $2\omega$, well beyond the gap at frequencies close to $E_{i}$ and hence much more sensitive to the basis set due to the involvement of high virtual states. On the other hand, as previously explained the difference between POC and REC is due to the damping and, as Figure \ref{FigBeta}d shows, both tensors are equal (besides discrete crossings) only in the low-frequency tail, where the absence of excitations makes the damping irrelevant. It can also be observed that $\Re$ POC changes sign at every local extrema of $\Im$ POC, in agreement with the Kramers-Kronig relations. The alignment need not be exact, and indeed the sign change does not realize in SHG (also fulfilling Kramers-Kronig) except at $E_{2}/2$, due to the significant contribution of multiple excitations. The behavior of REC is not restricted by these relations, instead it resembles the resonance structure of $\Re$ POC. Unlike $\Im\alpha$, all $\beta$ tensors are observed to change sign with the exception of $\Im$ SHG, which is only a feature of the two-photon resonances and not a general behavior at higher frequencies. 

The size of the flake has a noticeably larger effect on $\beta$ than on $\alpha$. The static value $\beta^{xxx}_{\text{STA}}$ grows from $-89.7$ to $-140.4$ (a.u.) between BN84 and BN282, whereas $\alpha^{xx}$ 
grows from 23.3 to 25.9 (a.u.), namely a change of $57\%$ versus $11\%$, all per B$-$N pair. In fact, the static values for the monolayer are $\alpha^{xx}=33.8$ and $\beta^{xxx}_{\text{STA}}=-339.3$ (a.u.) per unit cell, significantly far from the BN282 values. 
Similarly, the largest values of $|\beta^{xxx}_{\text{POC}}|$ and $|\beta^{xxx}_{\text{SHG}}|$ grow by factors of 2.2 and 1.7, against 1.3 for the maximum of $\Im\alpha^{xx}$. 
The resonances tend to red-shift and become better defined with size, e.g. the $E_1$ feature in $\Im$ POC, broad and weak in BN84, becomes well defined and red-shifted in the larger flakes. The largest observed change is in $\Re\beta^{xxx}_{\text{REC}}$ above $E_2$, which turns positive above $6.8$ eV for BN84 but stays negative up to 7.5 eV for BN282, reaching double the maximum absolute value. Therefore, at the BN282 size with the range-separated hybrid functional (including the full Fock-exchange at long-range) it is most likely that finite size effects impact $\beta$ significantly, in both magnitude and shape, and possibly also $\alpha$. These effects could be comparable to those of the functional itself, particularly for the hyperpolarizability.

\begin{figure}[H]
\centering 
\includegraphics[width=1.0\textwidth]{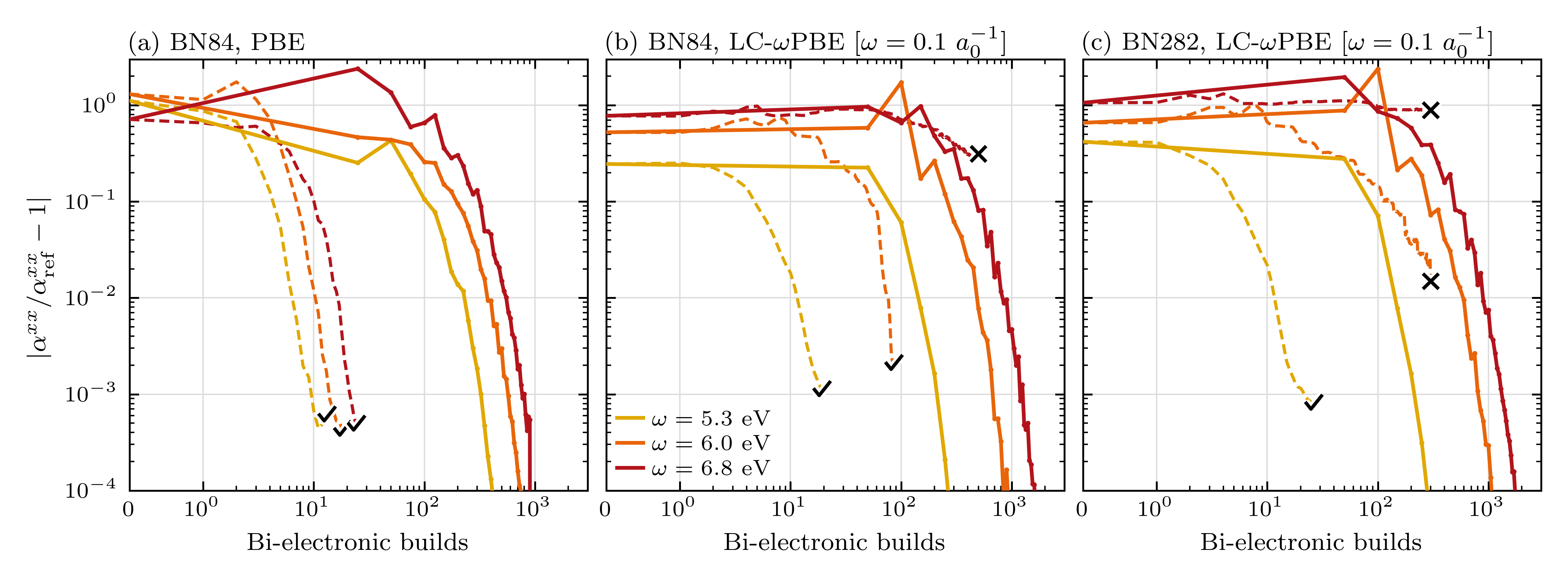}
\caption{\label{FigConv} Convergence of $\alpha^{xx}$ with the number of bi-electronic builds at $\omega=5.3$, $6.0$ and $6.8$ eV ($\eta=0.15$ eV damping) for (a) BN84 with PBE, (b) BN84 with LC-$\omega$PBE and (c) BN282 with LC-$\omega$PBE. Solid lines: Lanczos solver, resolved every 25 iterations (25 builds for PBE, 50 builds for LC-$\omega$PBE); their converged values define the reference $\alpha^{xx}_{\text{ref}}$. Dashed lines: self-consistent solver (DIIS), resolved every iteration (1 build); ending in a tick if converged or in a cross if stopped before convergence. The same Lanczos chain covers all frequencies, whereas for DIIS a different self-consistency is applied at each frequency. 
}
\end{figure}

\textbf{Lanczos vs self-consistent solver.}
The coupled-perturbed results in the hBN flakes were obtained with the Lanczos solver, and would have been unfeasible with the DIIS self-consistent solver when employing hybrid functionals. This is illustrated in Figure \ref{FigConv}, which displays the convergence evolution of $\alpha^{xx}$ with respect to the number of bi-electronic builds (for DIIS, one per iteration; for Lanczos, one per iteration for pure functionals and two for HF and hybrids) at selected frequencies, both below and beyond the first resonance, in BN84 and BN282. The reference $\alpha^{xx}_{\text{ref}}$ is the final converged value in the Lanczos method. In the DIIS method, the convergence criterion was set to an absolute change of $\alpha^{uu}$ below $0.5$ (a.u.) for 3 consecutive measurements (at every cycle). With this method, the combination of a relatively high absolute variation of $\alpha^{uu}$ with the requirement that it holds during a few iterations (not just once) was observed to be overall more satisfactory (avoiding false convergences and  excess iterations for a negligible final change) than looking for a one-time, tighter variation in large systems like these. A large history of 100 error vectors was kept in DIIS, which was observed beneficial with respect to smaller amounts. 

The impact of the functional is apparent between Figures \ref{FigConv}a and \ref{FigConv}b. With PBE, DIIS (dashed lines) reaches a relative error of $10^{-3}$ in 10$-$22 builds at the three frequencies, against 350$-$825 builds in the Lanczos method (solid lines). However, the latter delivers the whole spectrum (161 frequencies) for BN84, hence becoming the more efficient option for a spectrum of a few tens of frequencies or more. With the hybrid LC-$\omega$PBE, DIIS still converges below the first resonance $E_{1}=5.77$ eV (19 iterations at 5.3 eV), but the convergence is noticeable slower beyond it (85 iterations at 6.0 eV) and eventually stalls (not converged at 6.8 eV within 500 iterations). 
The most likely reason for this behavior lies in the self-consistent update, whose approximation to the excitation energies is built only from the energy differences $\varepsilon_a-\varepsilon_i$ in the denominators, see \eqref{U1}. With PBE, the excitations lie at or above the HOMO$-$LUMO gap, close to those differences. With LC-$\omega$PBE (and likewise for other hybrid functionals), some excited states lie within the gap, in this case with a large  $E_{g}-E_{1}=2.0$ eV (see Table \ref{TableE_hBN}). The difference between the excitation energy $\Omega_{I}$ and the contributing single-particle transitions $\varepsilon_a-\varepsilon_i$ can then be large, making some denominators $\varepsilon_a-\varepsilon_i-\omega>0$ and the target $\Omega_{I}-\omega<0$ have opposite signs (the number of them increasing with $\omega$), which only gets amplified in the self-consistent update. The only mechanism to correct this error is the DIIS extrapolation, which proves to be not enough beyond a certain frequency. The Lanczos method does not suffer from this problem, since it approximates the response by repeated application of the exact matrices $\mathbb{A}\pm\mathbb{B}$ to the dipole vector, independently of $\omega$, instead of building the excitation energies from the one-electron energy differences. 

The number of iterations for convergence of the Lanczos chain is largely insensitive to the flake size, as observed in Figures \ref{FigConv}b and \ref{FigConv}c. Reaching a relative error of $10^{-3}$ takes 250, 700, 1300 builds at 5.3, 6.0 and 6.8 eV for BN84, and 250, 850, 1350 builds for BN282. In contrast, the DIIS convergence deteriorates with system size above the first resonance. At 6.0 eV it converges for BN84 after 85 iterations, but for BN282 it still carries a relative error of $1.5\%$ at 300 iterations (which is not fully trustworthy in the absence of a reference). At 6.8 eV, deep into the continuum of excitations, the self-consistent method fails spectacularly, with a relative error for BN84 still at $31\%$ but slowly decreasing at 500 iterations, and a completely stalled convergence for BN282 after 300 iterations. This is consistent with the previous explanation for the DIIS failure, since the density of excitations beyond the absorption onset increases with system size.

\begin{figure}[H]
\centering 
\includegraphics[width=1.0\textwidth]{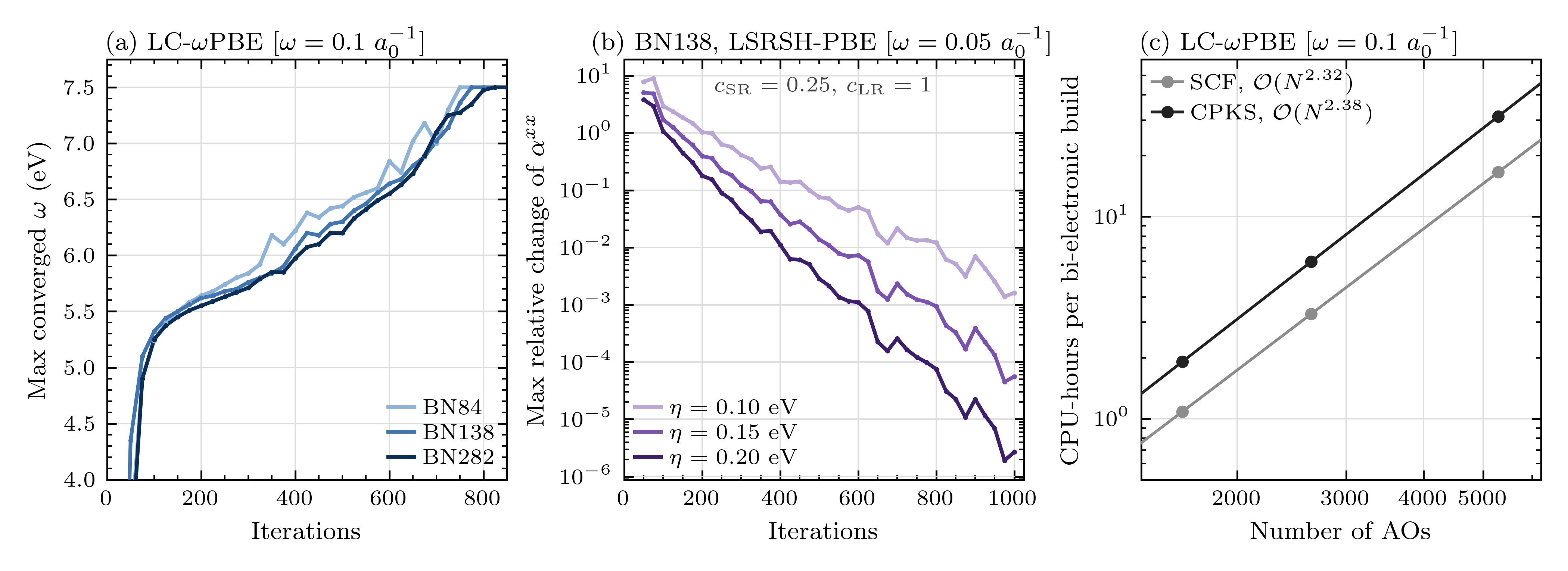}
\caption{\label{FigConv2} Convergence and cost of the Lanczos method.
(a) Highest frequency up to which $\alpha^{xx}$ is converged (relative change below $10^{-3}$) versus the number of Lanczos iterations, for the three hBN flakes with LC-$\omega$PBE and $\eta=0.15$ eV. Frequencies above 7.5 eV are not sampled. 
(b) Largest relative change of $\alpha^{xx}$ in BN138 with LSRSH-PBE over a [0.5, 8.5] eV grid, between checkpoints 25 iterations apart, for different damping factors.
(c) CPU-hours per bi-electronic build for the three hBN flakes with LC-$\omega$PBE and $\eta=0.15$ eV, employing 128 cores (within the ideal parallel-speedup region in all cases). Note that the numbers are highly sensitive to the \texttt{TOLINTEG} parameters, which were chosen for tight convergence of the bi-electronic integrals. Measured for both the hybrid ground-state and coupled-perturbed calculations, and fitted to a power law in the number of AOs. }
\end{figure}

\textbf{Convergence and scaling.}
To conclude we discuss other convergence and cost aspects of the Lanczos calculations in the hBN flakes, illustrated in Figure \ref{FigConv2} with hybrid functionals. Figure \ref{FigConv2}a shows that the required number of iterations for convergence increases with the frequency, though in a controlled way, in agreement with the results in Refs. \cite{Gruning2011_pseudoHermitianLanczos,Malcioglu2011_turboTDDFT}.
The response up to 5.5 eV, just below the first excitation, converges rapidly in up to 175 iterations, and beyond that it can be fitted to a very smooth parabola up to 7.5 eV (the maximum sampled frequency). Remarkably, the number of iterations at any frequency is essentially independent of system size, extrapolating the behavior of Figures \ref{FigConv}b and \ref{FigConv}c to the whole frequency range. The damping factor has a stronger effect on the convergence, as depicted in Figure \ref{FigConv2}b for the medium-size flake with the LSRSH-PBE functional. The largest relative change of $\alpha^{xx}$ across the frequency grid decreases an order of magnitude roughly every 275, 205, 165 iterations, respectively,
for $\eta=0.10$, $0.15$ and $0.20$ eV. Therefore, a finer spectral resolution generally requires longer Lanczos chains, although their cost in unchanged upon addition of other damping factors larger than the smallest one. 

The computational cost of the bi-electronic builds is shown in Figure \ref{FigConv2}c, for both the SCF and coupled-perturbed calculations with the range-separated hybrid LC-$\omega$PBE. Specifically, this comprises the calculation of the 4-center integrals for Hartree and Fock-exchange, and their contraction with the corresponding density matrix, see \eqref{B1} and \eqref{ApB}. The data were obtained with 128 cores (AMD EPYC 7H12@2.6GHz), a count which was observed to be in the ideal parallel-speedup region in all cases, and the wall-time is dominated by the calculation of the Fock exchange. The scaling with system size is essentially equal in both stages of the calculation, $\sim\mathcal{O}(N^{2.35})$. However, the coupled-perturbed build is $\sim1.8$ times more expensive than the SCF build. This is larger than the $\sim1.3$ increase from the non-hermiticity and complex nature of the density matrix, the reason being the exact evaluation of the Coulomb interactions in the coupled-perturbed calculation, see discussion below \eqref{dens_symmetrized}, and the reduced point group symmetry in the density matrix.
Since the number of iterations is nearly independent of the size, the scaling of a Lanczos calculation with system size is that of a ground-state hybrid calculation, though of course with a significantly larger prefactor which grows roughly linearly with the maximum frequency and the inverse of the damping factor.

\section{Conclusions}\label{secConclusions}
The damped, coupled-perturbed method for finite systems here described has been implemented in a developer version of \texttt{CRYSTAL} (version CRYSTAL23), and will be available in the next release of the code, currently in preparation. The implementation is compatible with many of the functionals available in the code, including global and range-separated hybrids based on LDA and GGA, in both closed-shell and collinear open-shell spin configurations. The first hyperpolarizability is obtained from first-order quantities through the $(2n+1)-$rule, involving a single damping factor (introduced as a relaxation mechanism in the TDGKS equations). 

The first-order equations are solved in the parallel-transport gauge with two methods that share the AO bi-electronic architecture, which exploits the full point group of the unperturbed system. One method is a self-consistent solver accelerated by DIIS, which requires a separate resolution for every frequency, damping factor and field direction. The other method is a Lanczos solver (symmetric product form), which yields the polarizability at any frequency and damping factor from a single chain (set of parameters) per field direction, as well as the first-order amplitudes required by $\beta$ at the cost of one additional bi-electronic build per frequency and direction. In practice, the Lanczos method is the preferred choice for any relatively dense frequency sampling (including the option of multiple damping factors) since its cost depends on the maximum frequency but not on the number of them. In fact, it is also the preferred choice with hybrid functionals beyond the absorption onset (for medium or large systems, even for single-frequency calculations), at which point the convergence of the self-consistent method is significantly slowed or stalled. The reason is the error amplification in approximating excitations below $\omega$ for which the electron-hole attraction is significant.
The drawbacks are that the Lanczos implementation requires setting \texttt{TOLINTEG}(4)$=$\texttt{TOLINTEG}(5) for hybrid functionals, slightly increasing the cost of the bi-electronic builds, and two of these builds per iteration instead of one.

The implementation was validated in smaller molecules up to 12 atoms against other codes \cite{Aidas2014_Dalton,NWCHEM2011} with excellent agreement in all cases, from which we selected the results in Figure \ref{FigH2O} as a representative example. As a case study, we have selected a series of three hBN flakes of up to 324 atoms and 5286 AOs. 
With semi-local functionals the absorption lacks the strongly bound first excitation of the monolayer, which is recovered with range-separated hybrids including the full Fock-exchange at long range.
LC-$\omega$PBE reproduces the position of the first peak of the periodic BSE spectrum but not the relative intensity of the second one, whereas a trade-off in the opposite direction is found when a short-range fraction of Fock-exchange is added (LSRSH-PBE). The first hyperpolarizability displays resonances at the same excitations, with weights that differ significantly from those in the absorption, and a much stronger dependence on the flake size. 
These indicate that finite-size effects are important at this scale in the nonlinear (and also linear) optical response. 
Admittedly, these proto-crystalline structures may not be the best candidates to show the accuracy of (adiabatic) TDDFT, unlike many regular molecules, nevertheless they provide a convenient platform to test the scalability of the implementation, while displaying interesting physics and showing the qualitative impact of hybrid functionals, as well as the quantitative differences between their different types.

In any case, we demonstrated that the Lanczos implementation is amenable for first and second-order hybrid calculations in large systems, with a scaling in the size that equals that of the ground-state calculation. An extension to periodic systems (including the multipolar treatment of the Coulomb interaction at long range), as well as an extension to meta-GGA functionals, are planned for future works.

\section*{Supporting Information}
Supporting Information available:  \texttt{CRYSTAL} input files for the H$_{2}$O and hBN flakes calculations. 

\appendix
\renewcommand{\thesection}{App \Alph{section}}
\section{The $(2n+1)-$rule with complex damping}\label{AppA}
\subsection{Elimination of the second-order coefficients}\label{AppA1}
Consider the first and second-order TDGKS without contraction with $c^{(0)\dagger}$, 
\begin{equation}\label{TDKS1_pre}
H^{u}(\omega)c^{(0)}+H^{(0)}c^{u}(\omega)-(\omega+i\eta)Sc^{u}(\omega)
-Sc^{u}(\omega)\varepsilon^{(0)}-Sc^{(0)}\lambda^{u}(\omega) = 0
\end{equation}
\begin{equation}\label{TDKS2_pre}\begin{aligned}
&H^{u,v}(\omega,\omega')c^{(0)} + H^{u}(\omega)c^{v}(\omega') + H^{v}(\omega')c^{u}(\omega) + H^{(0)}c^{u,v}(\omega,\omega') -  (\omega+\omega'+i\eta)Sc^{u,v}(\omega,\omega') 
\\
&- Sc^{u,v}(\omega,\omega')\varepsilon^{(0)} - Sc^{u}(\omega)\lambda^{v}(\omega') - Sc^{v}(\omega')\lambda^{u}(\omega)-Sc^{(0)}\lambda^{u,v}(\omega,\omega') = 0
\end{aligned}\end{equation}
and a general damping term with the form in \eqref{TDKS2_pre},  $-(\omega+\omega'+i\eta_{2})Sc^{u,v}(\omega,\omega')$. To illustrate the discussion in \ref{AppA2} we keep a general $\eta_{2}$, but note that for our purpose here $\eta_{2}=\eta$, the same damping factor as in the relaxation term in \eqref{TDKS}. Consider
\begin{equation}\label{beta_step1}
c^{u}(\omega+\omega')^{T} \cdot \left[\text{LHS of }\eqref{TDKS2_pre}\text{ for }v,w,\omega,\omega'\right] - \left[\text{LHS of }\eqref{TDKS1_pre}\text{ for }u,\omega+\omega'\right]^{T}\cdot c^{v,w}(\omega,\omega') = 0
\end{equation}
where $T$ denotes the transpose (no complex conjugation), and LHS alludes to the left-hand side of the respective equations. This leaves a term
\begin{equation}\label{A1}
A_{1}=U^{u}(\omega+\omega')^{T}\left[i(\eta-\eta_{2})U^{v,w}(\omega,\omega')-\lambda^{v,w}(\omega,\omega')\right]
\end{equation}
on the LHS of \eqref{beta_step1}. Then consider the equality
\begin{equation}\label{beta_step2}
\left[\text{LHS of }\eqref{TDKS2_pre}\text{ for }v,w,-\omega,-\omega'\right]^{\dagger}\cdot c^{u}(-\omega-\omega')^{*} -  c^{v,w}(-\omega,-\omega')^{\dagger} \cdot \left[\text{LHS of }\eqref{TDKS1_pre}\text{ for }u,-\omega-\omega'\right]^{*} = 0
\end{equation}
which leaves a term
\begin{equation}\label{A2}
A_{2}=\left[i(\eta_{2}-\eta)U^{v,w}(-\omega,-\omega')^{\dagger}-\lambda^{v,w}(-\omega,-\omega')^{\dagger}\right]U^{u}(-\omega-\omega')^{*}
\end{equation}
on the LHS of \eqref{beta_step2}. Now take
\begin{equation}
\Tr\left\{N\left[
\text{LHS of }\eqref{beta_step1} \:+\: \text{LHS of }\eqref{beta_step2}
\right]\right\}  =0
\end{equation}
and use \eqref{P1}, \eqref{P2}, \eqref{whermit}, \eqref{Uorth1}, \eqref{Uorth2}, and the choice of $\lambda^{u}(\omega)$ that leads to \eqref{U1}, to simplify the equation. Note that the arising term
\begin{equation}\label{core2nd}
\Tr\left[
P^{u}(-\omega-\omega')^{*}H^{v,w}(\omega,\omega')-P^{v,w}(\omega,\omega')H^{u}(-\omega-\omega')^{*}
\right] =
\Tr\left[\mu^{u}P^{v,w}(\omega,\omega')\right] +
\beta_{\text{xc}}^{uvw}(\omega,\omega')
\end{equation}
follows from \eqref{n1}, \eqref{n2}, \eqref{H1}, \eqref{H2} and \eqref{betaDFT}: the Hartree, exchange and first-order XC contributions to the two traces cancel by the permutational symmetry of the two-electron integrals and of the XC kernel [$f_{\text{xc}}^{\sigma,\sigma'}(\bm{r},\bm{r}')=f_{\text{xc}}^{\sigma',\sigma}(\bm{r}',\bm{r})$], and the first term on the right is \eqref{beta_v0}. The remaining terms are all of first order, and collecting them yields
\begin{equation}\label{Abeta}
\beta^{uvw}(\omega,\omega')=\left[\text{right-hand side of }\eqref{beta}\right]-\Tr\left\{N\left(A_{1}+A_{2}\right)\right\}
\end{equation}
The diagonal blocks of $\lambda^{v,w}$ do not contribute to $\Tr\{N(A_{1}+A_{2})\}$, since $U^{u}$ has vanishing diagonal blocks with our choice of $\lambda^{u}$. With block-diagonal multipliers, \eqref{A1}, \eqref{A2} and \eqref{Uorth2} then give
\begin{equation}\label{Aleft}
\Tr\left\{N\left(A_{1}+A_{2}\right)\right\}=i(\eta_{2}-\eta)\,\Tr\left\{N\left[U^{u}(-\omega-\omega')^{*},U^{v,w}(\omega,\omega')\right]_{-}\right\}
\end{equation}
which vanishes for our $\eta_{2}=\eta$ case. This proves \eqref{beta}, an expression for $\beta^{uvw}(\omega,\omega')$ that does not involve the second-order solution.

\subsection{Damping through the field: differences in the non-linear regime}\label{AppA2}
An alternative to introduce the damping parameter, commonly found in the literature \cite{Kummel2018,Giustino2018,Luppi2010_SHG}, is to include it in the external field instead of as a relaxation mechanism in the TDGKS equations \eqref{TDKS}. This is done by means of an adiabatic switching from $t\to-\infty$ \cite{TDPT1972}, whereby the field is expressed as $\bm{E}_{\eta}(t)=e^{\eta t}\bm{E}(t)$, with $\bm{E}(t)$ as in \eqref{Efield}. Every Fourier component of the field then carries the complex frequency $\omega+i\eta$, and the $n^{\text{th}}-$order terms in \eqref{Cexpans} the extra factor $e^{n\eta t}$. At first order, the time derivative of $e^{-i(\omega+i\eta)t}$ yields $-(\omega+i\eta)Sc^{u}(\omega)$, the same term as the relaxation in \eqref{TDKS1_pre}, so that \eqref{U1} and hence $\alpha^{uv}(\omega)$ are formally identical in both approaches. However, at second order the derivative introduces the term $-(\omega+\omega'+i2\eta)Sc^{u,v}(\omega,\omega')$, which introduces an extra factor of 2 to the damping with respect to \eqref{TDKS2_pre}. Thus, the derivation in \ref{AppA1} would apply with $\eta_{2}=2\eta$, hence \eqref{Aleft} becomes 
\begin{equation}\label{Aleft2}
\Tr\left\{N\left(A_{1}+A_{2}\right)\right\}=i\eta\,\Tr\left\{N\left[U^{u}(-\omega-\omega')^{*},U^{v,w}(\omega,\omega')\right]_{-}\right\}\neq0
\end{equation}
Therefore, the second-order solution cannot be removed from $\beta^{uvw}$ in this approach if the first-order equations are taken at $\eta$. Instead, \eqref{Aleft2} vanishes only if those are taken at $2\eta$, which requires the resolution of the first-order equations at two different damping factors (trivial with the Lanczos solver, twice as expensive with DIIS). 
Furthermore, the full second-order solution at any frequencies would carry twice the damping of the relaxation approach. This can be seen from the expression of $U^{u,v}$, whose off-diagonal blocks are fixed by \eqref{TDKS2_pre} and its diagonal blocks only determined up to a gauge choice by \eqref{Uorth2}, which can be chosen e.g. as
\begin{equation}\label{U2}
U^{u,v}_{m,n}(\omega,\omega')=\left\{\begin{aligned}
&\frac{G^{u,v}_{m,n}(\omega,\omega')+T^{u,v}_{m,n}(\omega,\omega')}{\omega+\omega'+\varepsilon^{(0)}_{n}-\varepsilon^{(0)}_{m}+i\eta} , \;\;
&m\:\text{occ},n\:\text{virt} ; \text{ or } m\:\text{virt},n\:\text{occ} 
\\
& \frac{1}{2}[U^{u}(\omega)U^{v}(\omega')+
U^{v}(\omega')U^{u}(\omega)]_{m,n}, \;\;
&m\:\text{occ},n\:\text{occ} ; \text{ or } m\:\text{virt},n\:\text{virt} 
\end{aligned}\right.
\end{equation}
\begin{equation}
T^{u,v}_{m,n}(\omega,\omega')=
\left[
G^{u}(\omega)U^{v}(\omega') +  G^{v}(\omega')U^{u}(\omega) 
- U^{u}(\omega)G^{v}(\omega') - U^{v}(\omega')G^{u}(\omega)
\right]_{m,n}
\end{equation}
Correspondingly, in the damped-field route the denominator of the off-diagonal blocks of \eqref{U2} becomes $\omega+\omega'+\varepsilon^{(0)}_{n}- \varepsilon^{(0)}_{m}+i2\eta$. This behavior extends to higher orders through the corresponding damping multiples.

As an observation, it could be tempting to try canceling the term \eqref{Aleft2} by choosing off-diagonal second-order multipliers, specifically $\lambda^{u,v}_{m,n}(\omega,\omega')=-i\eta U^{u,v}_{m,n}(\omega,\omega')$ for 
$(m,n)\in(\text{occ,virt})$ and $(\text{virt,occ})$, which indeed makes $\Tr\{NA_{1}\}$ and $\Tr\{NA_{2}\}$ vanish separately and is compatible with
\eqref{lambdaC} at second order due to \eqref{Uorth2}. However, as argued in the main text this is not admissible due to the arbitrary contribution that the virt-occ block would introduce in the evolution of the density matrix.

\section{Exploitation of symmetry}\label{AppB}
The AO index can be subdivided as $\mu=(A,\lambda,l,m)$, where $A$ labels the atoms, $\lambda=\lambda(A)$ labels the shells for the atom, and $l=l(\lambda)$, $m=m(l)$ label the orbital angular momentum quantum numbers within the shell. Under the action of a spatial (unitary) symmetry operation $g$ of the molecular point group $\mathcal{G}$, the AOs transform as
\begin{equation}\label{transform}
\varphi_{(A,\lambda,l,m)}(g\bm{r})=\sum_{m'=-l}^{l}\mathcal{D}^{l}_{m,m'}(g)\varphi_{(g^{-1}A,\lambda,l,m')}(\bm{r})
\end{equation}
where $\mathcal{D}^{l}$ is the representation of angular momentum $l$ in the basis of real-valued solid spherical harmonics, of dimension $2l+1$ although in general reducible, and $g^{-1}A$ is the atom obtained from $A$ by applying $g^{-1}$.   

The application of the electric field along $u$ does not alter the AOs but reduces the point group $\mathcal{G}$ to a subgroup $\mathcal{G}_{u}$ consisting of all the operations that leave the direction $u$ invariant (e.g. a rotation along the $u$ axis), under which the perturbed density matrix of that direction is invariant. To clarify, we are not considering spin groups here, hence each operation acts simultaneously on coordinate and spin spaces (the inversion acting as the identity in the latter). For an original symmetry $g\in\mathcal{G}$, consider the bi-electronic term $B^{\sigma\;u}_{(gA_{\mu},\lambda_{\mu},l_{\mu},m_{\mu}),(gA_{\nu},\lambda_{\nu},l_{\nu},m_{\nu})}(\omega)$ from \eqref{B1}. Focusing only on the Fock-exchange (the reasoning for the Hartree term is analogous) and dropping the shell indices as they play a trivial role regarding transformations, we can write
\begin{equation}\label{trans_v1}\begin{aligned}
&\sum_{\rho,\tau}P^{\sigma\;u}_{\rho,\tau}(\omega)\int\varphi_{(gA_{\mu},l_{\mu},m_{\mu})}(\bm{r})\varphi_{\rho}(\bm{r})\frac{1}{\abs{\bm{r}-\bm{r}'}}\varphi_{(gA_{\nu},l_{\nu},m_{\nu})}(\bm{r}')\varphi_{\tau}(\bm{r}')d\bm{r}d\bm{r}' = \\
&\sum_{\rho,\tau}P^{\sigma\;u}_{\rho,\tau}(\omega)\int\varphi_{(gA_{\mu},l_{\mu},m_{\mu})}(g\bm{r})\varphi_{\rho}(g\bm{r})\frac{1}{\abs{\bm{r}-\bm{r}'}}\varphi_{(gA_{\nu},l_{\nu},m_{\nu})}(g\bm{r}')\varphi_{\tau}(g\bm{r}')d\bm{r}d\bm{r}' = \\
&\sum_{\rho,\tau}P^{\sigma\;u}_{\rho,\tau}(\omega)\sum_{m'_{\mu},m'_{\nu}}\mathcal{D}^{l_{\mu}}_{m_{\mu},m'_{\mu}}(g)\mathcal{D}^{l_{\nu}}_{m_{\nu},m'_{\nu}}(g)\int\varphi_{(A_{\mu},l_{\mu},m'_{\mu})}(\bm{r})\varphi_{\rho}(g\bm{r})\frac{1}{\abs{\bm{r}-\bm{r}'}}\varphi_{(A_{\nu},l_{\nu},m'_{\nu})}(\bm{r}')\varphi_{\tau}(g\bm{r}')d\bm{r}d\bm{r}' 
\end{aligned}\end{equation}
where we have first applied an orthogonal change of coordinates within the integral, and then used \eqref{transform}. If $g\in\mathcal{G}_{u}$ then, using the invariance of the perturbed density matrix
under a symmetry of the perturbed system, 
\begin{equation}\label{symm1B}
B^{\sigma\;u}_{(gA_{\mu},l_{\mu},m_{\mu}),(gA_{\nu},l_{\nu},m_{\nu})}(\omega) = 
\sum_{m'_{\mu},m'_{\nu}}\mathcal{D}^{l_{\mu}}_{m_{\mu},m'_{\mu}}(g)\mathcal{D}^{l_{\nu}}_{m_{\nu},m'_{\nu}}(g)
B^{\sigma\;u}_{(A_{\mu},l_{\mu},m'_{\mu}),(A_{\nu},l_{\nu},m'_{\nu})}(\omega)
\end{equation}
as expected. If $g\notin\mathcal{G}_{u}$, that is, if $g$ is no longer a symmetry under the field along $u$, we can continue with \eqref{trans_v1} by applying \eqref{transform} to get
\begin{equation}\label{trans_v2}\begin{aligned}
&\sum_{m'_{\mu},m'_{\nu}}\mathcal{D}^{l_{\mu}}_{m_{\mu},m'_{\mu}}(g)\mathcal{D}^{l_{\nu}}_{m_{\nu},m'_{\nu}}(g)
\sum_{\rho,\tau}P^{\sigma\;u}_{(gA_{\rho},l_{\rho},m_{\rho}),(gA_{\tau},l_{\tau},m_{\tau})}(\omega) 
\sum_{m'_{\rho},m'_{\tau}}\mathcal{D}^{l_{\rho}}_{m_{\rho},m'_{\rho}}(g)\mathcal{D}^{l_{\tau}}_{m_{\tau},m'_{\tau}}(g) \cdot \\
&\int\varphi_{(A_{\mu},l_{\mu},m'_{\mu})}(\bm{r})\varphi_{(A_{\rho},l_{\rho},m'_{\rho})}(\bm{r})\frac{1}{\abs{\bm{r}-\bm{r}'}}\varphi_{(A_{\nu},l_{\nu},m'_{\nu})}(\bm{r}')\varphi_{(A_{\tau},l_{\tau},m'_{\tau})}(\bm{r}')d\bm{r}d\bm{r}' 
\end{aligned}\end{equation}
Introducing 
\begin{equation}
\tilde{P}^{\sigma\;u}_{\rho,\tau}(g,\omega)=
\sum_{m'_{\rho},m'_{\tau}}P^{\sigma\;u}_{(gA_{\rho},l_{\rho},m'_{\rho}),(gA_{\tau},l_{\tau},m'_{\tau})}(\omega)
\mathcal{D}^{l_{\rho}}_{m'_{\rho},m_{\rho}}(g)\mathcal{D}^{l_{\tau}}_{m'_{\tau},m_{\tau}}(g) \:,
\end{equation}
\begin{equation}
\tilde{B}^{\sigma\;u}_{\mu,\nu}(g,\omega)=
\sum_{\rho,\tau}\sum_{\sigma'}\tilde{P}^{\sigma'\;u}_{\rho,\tau}(g,\omega)
\left[(\mu\nu\vert\rho\tau) -\delta_{\sigma,\sigma'}(\mu\rho\vert\nu\tau)\right]
\end{equation}
we therefore obtain 
\begin{equation}\label{symm2B}
B^{\sigma\;u}_{(gA_{\mu},l_{\mu},m_{\mu}),(gA_{\nu},l_{\nu},m_{\nu})}(\omega) = 
\sum_{m'_{\mu},m'_{\nu}}\mathcal{D}^{l_{\mu}}_{m_{\mu},m'_{\mu}}(g)\mathcal{D}^{l_{\nu}}_{m_{\nu},m'_{\nu}}(g)
\tilde{B}^{\sigma\;u}_{(A_{\mu},l_{\mu},m'_{\mu}),(A_{\nu},l_{\nu},m'_{\nu})}(g,\omega)
\end{equation}
With \eqref{symm1B} and \eqref{symm2B} we can employ the whole unperturbed point group to generate linearly dependent bi-electronic matrix elements in terms of a minimal set of irreducible 4-center integrals. The difference between evaluating both, namely in using a symmetry that is removed or not, is not significant in terms of computational cost. The symmetry-based infrastructure was programmed in a previous implementation \cite{thesis_jacopo}, which we adapted to the present non-hermitian, damped formulation.

\section{Open-shell GGA expressions of the XC kernel up to $2^{\text{nd}}-$order}\label{AppC}
Mostly for the purpose of documentation, we indicate the expressions for the functional derivatives in terms of the XC energy density $\epsilon_{\text{xc}}$ required for an open-shell GGA response calculation up to second order. 
For the local spin density and generalized gradient approximations (LSDA and GGA) to collinear spin DFT, the XC energy is the integral of an energy density,
\begin{equation}\label{ExcGGAos}
E_{\text{xc}}[n_\uparrow,n_\downarrow]=\int\epsilon_{\text{xc}}
\left(n_{\uparrow}(\bm{r}),n_{\downarrow}(\bm{r}),\abs{\bm{\nabla}n_{\uparrow}(\bm{r})}^{2},\abs{\bm{\nabla}n_{\downarrow}(\bm{r})}^{2},\bm{\nabla}n_{\uparrow}(\bm{r})\cdot\bm{\nabla}n_{\downarrow}(\bm{r})\right)d\bm{r}
\end{equation}
We omit the frequency and position arguments for compactness, all of which are just $\bm{r}$ due to the absence of non-locality in $\epsilon_{\text{xc}}$. 
In the implementation, the partial derivatives of $\epsilon_{\text{xc}}$ up to third order are obtained with \texttt{XCFun} \cite{Ekstrom2010_xcfun}.

The zeroth order \eqref{XCmat0} reads (with $\gamma_{\sigma,\sigma'}=\bm{\nabla}n_{\sigma}\cdot\bm{\nabla}n_{\sigma^\prime}$ and $d_{\sigma,\sigma'}=1+\delta_{\sigma,\sigma'}$)
\begin{equation}\label{eqn:vxc0}
V^{\sigma\;(0)}_{\mu,\nu} =\int\frac{\partial\epsilon_{\text{xc}}}{\partial n_{\sigma}}\varphi_{\mu}\varphi_{\nu} d\bm{r} + \sum_{\sigma^\prime}d_{\sigma,\sigma'}\int  \frac{\partial\epsilon_{\text{xc}}}{\partial \gamma_{\sigma,\sigma'}}\bm{\nabla}n_{\sigma^\prime}
\cdot\bm{\nabla}(\varphi_{\mu}\varphi_{\nu})d\bm{r} 
\end{equation}
The derivative of Eq. \eqref{eqn:vxc0} provides the first-order XC matrix \eqref{V1}, namely (with $\gamma^u_{\sigma,\sigma'}=\bm{\nabla}n_{\sigma}\cdot\bm{\nabla}n_{\sigma^\prime}^u$ and omitting the frequency argument)
\begin{eqnarray}\label{eqn:vxc1}
V^{\sigma\;u}_{\mu,\nu} = \sum_{\sigma'} \int    \Bigg[ \frac{\partial^2\epsilon_{\text{xc}}}{\partial n_{\sigma} \partial n_{\sigma'} } n^u_{\sigma'}+ \sum_{\sigma''}d_{\sigma',\sigma''} \frac{\partial^2\epsilon_{\text{xc}}}{\partial n_{\sigma} \partial \gamma_{\sigma',\sigma''}  } \gamma^u_{\sigma',\sigma''} \Bigg] \varphi_{\mu} \varphi_{\nu} d\bm{r} + d_{\sigma,\sigma'} \Bigg[ \frac{\partial\epsilon_{\text{xc}}}{ \partial \gamma_{\sigma,\sigma'}} \bm{\nabla}n_{\sigma'}^u \nonumber \\
+\sum_{\sigma''}\Bigg(
\frac{\partial^2\epsilon_{\text{xc}}}{\partial n_{\sigma''} \partial \gamma_{\sigma,\sigma'}} n_{\sigma''}^u  + \sum_{\sigma'''} 
d_{\sigma'',\sigma'''}\frac{\partial^2\epsilon_{\text{xc}}}{\partial \gamma_{\sigma,\sigma'} \partial \gamma_{\sigma'',\sigma'''}} \gamma^u_{\sigma'',\sigma'''} \Bigg)
\bm{\nabla}n_{\sigma'} \Bigg] \cdot \bm{\nabla}(\varphi_{\mu} \varphi_{\nu}) d\bm{r}
\end{eqnarray}

At second order we are required to evaluate only $\beta^{uvw}_{\text{xc}}$ \eqref{betaDFT}, which involves the second-order XC kernel. Introducing $\gamma^{uv}_{\sigma,\sigma'}=\bm{\nabla}n_{\sigma}^{u}\cdot\bm{\nabla}n_{\sigma'}^{v}$ we can write

\begin{equation}\label{eqn:betaxc}\begin{aligned}
&\beta^{uvw}_{\text{xc}} = \sum_{\sigma_{0},\sigma_{1},\sigma_{2}} \int d\bm{r}   \frac{\partial^3\epsilon_{\text{xc}}}{\partial n_{\sigma_{0}} \partial n_{\sigma_{1}} \partial n_{\sigma_{2}} } n^u_{\sigma_{0}}n^v_{\sigma_{1}}n^w_{\sigma_{2}} + 
d_{\sigma_{0},\sigma_{1}}\frac{\partial^2\epsilon_{\text{xc}}}{\partial n_{\sigma_{2}}\partial\gamma_{\sigma_{0},\sigma_{1}}} ( n^u_{\sigma_{2}}\gamma^{vw}_{\sigma_{0},\sigma_{1}} + n^{v}_{\sigma_{2}}\gamma^{uw}_{\sigma_{0},\sigma_{1}}+  n^{w}_{\sigma_{2}}\gamma^{uv}_{\sigma_{0},\sigma_{1}}) 
 \\
&+
\sum_{\sigma_{3}}d_{\sigma_{0},\sigma_{1}}\Bigg\{ 
\frac{\partial^3\epsilon_{\text{xc}}}{\partial n_{\sigma_{2}} \partial n_{\sigma_{3}} \partial \gamma_{\sigma_{0},\sigma_{1}} } 
\left[ n^{v}_{\sigma_{2}}n^{w}_{\sigma_{3}}\gamma^{u}_{\sigma_{0},\sigma_{1}} + n^{u}_{\sigma_{2}}(n^{v}_{\sigma_{3}}\gamma^{w}_{\sigma_{0},\sigma_{1}} + n^{w}_{\sigma_{3}}\gamma^{v}_{\sigma_{0},\sigma_{1}}) \right] + 
d_{\sigma_{2},\sigma_{3}}\frac{\partial^2\epsilon_{\text{xc}}}{\partial \gamma_{\sigma_{0},\sigma_{1}}\partial\gamma_{\sigma_{2},\sigma_{3}}} %\cdot
\\
& 
( \gamma^{u}_{\sigma_{1},\sigma_{0}}\gamma^{vw}_{\sigma_{2},\sigma_{3}} + \gamma^{v}_{\sigma_{2},\sigma_{3}}\gamma^{uw}_{\sigma_{0},\sigma_{1}} + \gamma^{w}_{\sigma_{2},\sigma_{3}}\gamma^{uv}_{\sigma_{0},\sigma_{1}} ) + 
\sum_{\sigma_{4}}d_{\sigma_{2},\sigma_{3}} \Bigg[ \frac{\partial^3\epsilon_{\text{xc}}}{\partial n_{\sigma_{4}} \partial \gamma_{\sigma_{0},\sigma_{1}} \partial \gamma_{\sigma_{2},\sigma_{3}} } 
\left[ n^{u}_{\sigma_{4}}\gamma^{v}_{\sigma_{0},\sigma_{1}}\gamma^{w}_{\sigma_{2},\sigma_{3}} + \right.
 \\  &\left. 
(n^{v}_{\sigma_{4}}\gamma^{w}_{\sigma_{2},\sigma_{3}} + n^{w}_{\sigma_{4}}\gamma^{v}_{\sigma_{2},\sigma_{3}})\gamma^{u}_{\sigma_{0},\sigma_{1}} \right] +  
\sum_{\sigma_{5}} d_{\sigma_{4},\sigma_{5}}\frac{\partial^3\epsilon_{\text{xc}}}{\partial \gamma_{\sigma_{0},\sigma_{1}} \partial \gamma_{\sigma_{2},\sigma_{3}} \partial \gamma_{\sigma_{4},\sigma_{5}} } \gamma^{u}_{\sigma_{0},\sigma_{1}}\gamma^{v}_{\sigma_{2},\sigma_{3}}\gamma^{w}_{\sigma_{4},\sigma_{5}} \Bigg] \Bigg\}
\end{aligned}\end{equation}
It can indeed be observed that all terms are first-order, in consistency with the $(2n+1)-$rule. The implementation of Eq. \eqref{eqn:betaxc} is based on that reported in Ref. \cite{orlando2010calculation}.


\begin{acknowledgement}
A.G.B acknowledges financial support from Universidad
Aut\'onoma de Madrid through a FPI-UAM grant and the UAM-SANTANDER (2025) program for the mobility of young researchers, which enabled the start of this work.
A.G.B. thankfully acknowledges Red Española de Supercomputaci\'on for the computational resources provided by Universidad de M\'alaga  through the projects FI-2025-3-0026 and FI-2026-1-0018.
This research has received funding from the Project CH4.0 under the MUR program ``Dipartimenti di Eccellenza 2023-2027'' (CUP: D13C22003520001).
\end{acknowledgement}

The authors declare no competing financial interest.


%\appendix
%\section{Appendix A}
%A

%%%REFERENCES%%%

%\bibliographystyle{jabbrv}
\bibliography{main} 

\end{document}